\documentclass[%
reprint,
groupedaddress,
longbibliography,
nofootinbib,
amsmath,amssymb,
floatfix,
aip,
author-year,
]{revtex4-1}

\usepackage[utf8]{inputenc}
\usepackage{amsthm}
\usepackage{amsmath}
\usepackage{graphicx}
\usepackage{dcolumn}
\usepackage{bm}
\usepackage[table]{xcolor}
\usepackage{placeins}
\definecolor{Set1-blue}{RGB}{55,126,184}
\usepackage[colorlinks=true,%
            linkcolor   = {Set1-blue},%
            citecolor   = {Set1-blue},%
            urlcolor    = {Set1-blue}]{hyperref}

\usepackage{booktabs}
\AtBeginDocument{
\heavyrulewidth=.08em
\lightrulewidth=.05em
\cmidrulewidth=.03em
\belowrulesep=.65ex
\belowbottomsep=0pt
\aboverulesep=.4ex
\abovetopsep=0pt
\cmidrulesep=\doublerulesep
\cmidrulekern=.5em
\defaultaddspace=.5em
}
\usepackage{longtable}
\usepackage{xparse}

\usepackage{chngcntr}

\DeclareMathOperator{\PP}{PP}
\DeclareMathOperator{\MAD}{MAD}
\DeclareMathOperator{\MAPD}{MAPD}
\DeclareMathOperator{\median}{median}
\DeclareMathOperator{\Pub}{P}
\DeclareMathOperator{\lognormal}{LogNormal}

\graphicspath{ {figures/} }

\setcitestyle{authoryear,round}
\begin{document}

\title{Systematic analysis of agreement between metrics and peer review in the UK REF}

\author{V.A. Traag}
\email{v.a.traag@cwts.leidenuniv.nl}
\affiliation{Centre for Science and Technology Studies (CWTS), Leiden University, the Netherlands}

\author{L. Waltman}
\affiliation{Centre for Science and Technology Studies (CWTS), Leiden University, the Netherlands}

\date{\today}

\begin{abstract}
When performing a national research assessment, some countries rely on citation metrics whereas others, such as the UK, primarily use peer review.
In the influential \emph{Metric Tide} report, a low agreement between metrics and peer review in the UK Research Excellence Framework (REF) was found.
However, earlier studies observed much higher agreement between metrics and peer review in the REF and argued in favour of using metrics.
This shows that there is considerable ambiguity in the discussion on agreement between metrics and peer review.
We provide clarity in this discussion by considering four important points: (1) the level of aggregation of the analysis; (2) the use of either a size-dependent or a size-independent perspective; (3) the suitability of different measures of agreement; and (4) the uncertainty in peer review.
In the context of the REF, we argue that agreement between metrics and peer review should be assessed at the institutional level rather than at the publication level.
Both a size-dependent and a size-independent perspective are relevant in the REF.
The interpretation of correlations may be problematic and as an alternative we therefore use measures of agreement that are based on the absolute or relative differences between metrics and peer review.
To get an idea of the uncertainty in peer review, we rely on a model to bootstrap peer review outcomes.
We conclude that particularly in Physics, Clinical Medicine, and Public Health, metrics agree quite well with peer review and may offer an alternative to peer review.
\end{abstract}

\maketitle

\section{Introduction}

\noindent Many countries have some form of a national research assessment exercise in which universities and other research institutions are evaluated~\citep{Hicks2012}.
In part, such assessments aim to account for the expenses of public funds, but sometimes they also function to distribute funds based on performance.
Scientific quality or scientific impact plays a central role in many assessment exercises~\citep{Zacharewicz2018}, but institutions may also be evaluated on other performance dimensions, such as their societal, cultural, and economic impact.
Here, we restrict ourselves to scientific quality or scientific impact determined based on the publication output of an institution.
However, we acknowledge that other dimensions may also play a critical role.

How the quality or impact of publications is assessed differs from country to country.
Some countries have a national research assessment exercise that is driven by citation metrics, whereas others rely on peer review~\citep{Hicks2012}.
In particular, the United Kingdom (UK) has a long tradition of research assessment that relies on peer review, starting with the first assessment exercise in 1986.
The latest assessment exercise, referred to as the Research Excellence Framework (REF), took place in 2014.
It was followed by a detailed report, known as the \emph{Metric Tide} report~\citep{Wilsdon2015}, that critically examined the possible role of citation metrics in the REF.
It concluded that ``[m]etrics should support, not supplant, expert judgement''~\citep[p. viii]{Wilsdon2015}.
To support this conclusion, the report provided statistical evidence of the lack of agreement between metrics and peer review.
Here, we re-examine the statistical evidence for this conclusion.
The \emph{Metric Tide} report also offered other arguments to support the above conclusion.
It argued that metrics are contested among academics, and should therefore not be used, whereas peer review commands widespread support.
Moreover, metrics may create perverse incentives. 
We do not consider these arguments further in this paper, and restrict ourselves to the statistical argument presented in the \emph{Metric Tide}.
Of course, the other arguments should play a role in the broader discussion on the relative merits of peer review and metrics.

The various assessment exercises carried out in the UK during the past decades have all been accompanied by papers that compare citation metrics and peer review.
Although the results vary from field to field, most studies found correlations of about $0.7$.
Some authors obtained higher correlations, on the order of $0.9$.
However, the \emph{Metric Tide} report found significantly lower correlations in the range of about $0.2$--$0.4$.
Interestingly, even when authors obtained similar correlations, they did not always draw the same conclusion.
Some, such as \citet{Mryglod2015a} and \citet{Mahdi2008}, argued that a correlation of $0.7$ is too low to consider using metrics, while others, such as \citet{Thomas1998} and \citet{Taylor2011}, argued that a correlation of $0.7$ is sufficiently high.

We try to provide clarity in this debate by considering four important points:
\begin{enumerate}
  \item The agreement between metrics and peer review depends on the level of aggregation. The level of individual publications constitutes the lowest level of aggregation. The level of researchers and the level of research institutions represent higher levels of aggregation.
  \item At aggregate levels, metrics and peer review may take a size-dependent perspective---scaling with the size of an institution---or a size-independent perspective---being independent of the size of an institution. This distinction is particularly relevant when reporting correlations.
  \item Correlations between metrics and peer review may not be the most informative measure of agreement. Other measures may be more appropriate.
  \item Peer review is subject to uncertainty. This should be taken into consideration when interpreting the agreement between metrics and peer review.
\end{enumerate}

We first briefly discuss the REF and consider its objectives.
This is followed by a review of the literature on comparing metrics and peer review in the context of the REF and its precursors.
We argue that in the REF context, proper comparisons between metrics and peer review should be made at the institutional level, not at the level of individual publications.
We also briefly discuss how a size-dependent perspective relates to a size-independent perspective.
As we show, size-dependent correlations can be high even if the corresponding size-independent correlations are low.
We then introduce two measures of agreement that we consider to be more informative than correlations.
One measure is especially suitable for the size-dependent perspective, while the other measure is more suitable for the size-independent perspective.
To get some idea of the uncertainty in peer review, we introduce a simple model of peer review.

Based on our analysis, we conclude that for some fields, the agreement between metrics and peer review is similar to the internal agreement of peer review.
This is the case for three fields in particular: Clinical Medicine, Physics, and Public Health, Health Services \& Primary Care.
Finally, we discuss the implications of our findings for the REF 2021 that is currently in preparation.

\section{UK Research Excellence Framework}

\noindent The UK REF has three objectives:
\begin{quote}
  \begin{enumerate}
    \item To provide accountability for public investment in research and produce evidence of the benefits of this investment.
    \item To provide benchmarking information and establish reputational yardsticks, for use within the [Higher Education] sector and for public information.
    \item To inform the selective allocation of funding for research. 
  \end{enumerate}
  {\hfill \footnotesize From \url{http://www.ref.ac.uk/about/whatref/} for REF
    2021.\footnote{Interestingly, the order of these objectives for REF 2014
  are different, see \url{https://www.ref.ac.uk/2014/about/}.}}
\end{quote}
In addition, three further roles that the REF fulfills were identified:
\begin{quote}
  \begin{enumerate}
    \setcounter{enumi}{3}
    \item To provide a rich evidence base to inform strategic decisions about national research priorities.
    \item To create a strong performance incentive for HEIs and individual researchers.
    \item To inform decisions on resource allocation by individual HEIs and other bodies.
  \end{enumerate}
  {\hfill \footnotesize From \url{https://www.ref.ac.uk/media/1050/ref2017_01.pdf}}
\end{quote}

To meet these objectives, the REF assesses institutions in terms of (1) research output, (2) societal impact of the research, and (3) the environment supporting the research.
Here, we are concerned only with the assessment of research output.
In the REF 2014, the assessment of research output accounted for $65\%$ of the overall assessment of institutions.
Each output evaluated in the REF 2014 was awarded a certain number of stars: four stars indicates world-leading research, three stars indicates internationally excellent research, two stars indicates internationally recognised research, and one star indicates nationally recognised research.

The three above-stated objectives are each addressed in a different way.
The overall \emph{proportion} of high-quality research that has been produced is relevant for the first objective.
Indeed, the REF 2014 website boasts that $30\%$ of the submitted UK research was world-leading four-star research: public investment results in high-quality science.
The \emph{proportion} of research outputs awarded a certain number of stars also provides a reputational yardstick for institutions and thereby serves the second objective.
Indicators based on these proportions feature in various league tables constructed by news outlets such as the Guardian and Times Higher Education.
Such indicators may influence the choice of students and researchers regarding where to study and perform research. 
The \emph{total number} of publications that were awarded four or three stars influences the distribution of funding, which is relevant for the third objective of the REF.

Hence, the objective of establishing a reputational yardstick corresponds to a size-independent perspective, while the objective of funding allocation corresponds to a size-dependent perspective.
This means that agreement between metrics and peer review is relevant from both perspectives.
We will comment in more detail on the distinction between the two perspectives in Section~\ref{sec:size_dependent}.

To provide an indication of the importance of the REF 2014, we briefly look at the funding of UK higher education in 2017--2018\footnote{\url{http://www.hefce.ac.uk/funding/annallocns/1718/}}. 
In 2017--2018, REF results based on research output were used by the Higher Education Funding Council for England (HEFCE) to allocate \pounds$685$M to institutions.
Although many details are involved (e.g. extra funding for the London region, weighing cost-intensive fields), this was based largely on $4^*$ and $3^*$ publications, which were awarded roughly $80\%$ and $20\%$ of the money, respectively.
This amounted to about \pounds$10\,000$ per $4^*$ publication and about \pounds$2\,000$ per $3^*$ publication per year on average\footnote{In the REF 2014, in total $42\,481$ publications were awarded $4^*$ and $94\,153$ publications were awarded $3^*$. In reality, calculations are more complex, as they involve the number of staff in FTE, subject cost weights, and specific weights for the London area.}.
The total amount of about \pounds$685$M allocated through the evaluation of research output represented about $20\%$ of the total budget of HEFCE of \pounds$3\,602$M and about $40\%$ of the total research budget of HEFCE of \pounds$1\,606$M.
As such, it is a sizeable proportion of the total budget.

\section{Literature Review}

\noindent We review previous literature on how metrics compare to peer review in previous research assessment exercises in the UK.
We then briefly review literature that analyses how metrics and peer review compared in the REF 2014.

\subsection{Research Assessment Exercise}

\noindent In 1986, the University Grants Committee (UGC) undertook the first nationwide assessment of universities in the UK, called the research selectivity exercise.
Its primary objective was to establish a more transparent way of allocating funding, especially in the face of budget cuts~\citep{Jump2014}. 
Only two years later, \citet{Crewe1988} undertook the first bibliometric comparison of the results for Politics departments in the first 1986 exercise. 
The results of the 1986 exercise were announced per cost centre (resembling somewhat a discipline or field) of a university in terms of four categories: outstanding, above average, about average, and below average.
This limited the possibilities for bibliometric analysis somewhat, and~\citet{Crewe1988} only made some basic comparisons based on the number of publications.
He concluded that ``there is a close but far from perfect relationship between the UGC's assessment and rankings based on publication records''\citep[p. 246]{Crewe1988}.
Indeed, later exercises also showed that higher ranked institutions are typically larger (in terms of either staff or publications).
In the same year, \citet{Carpenter1988} analysed Physics and Chemistry outcomes of the UGC exercise.
They compared the outcomes to a total influence score, a type of metric similar to the Eigenfactor~\citep{Bergstrom2007}, and found a correlation of $0.63$ for Physics and $0.77$ for Chemistry.
The total influence score used by~\citet{Carpenter1988} is size-dependent, and the average influence per paper showed a correlation of only $0.22$ and $0.34$ for Physics and Chemistry, respectively.
It is not clear whether the 1986 UGC results themselves were size-dependent or size-independent.

The next research selectivity exercise in 1989 was undertaken by the Universities Funding Council (UFC).
The exercise made some changes and allowed universities to submit up to two publications per research staff~\citep{Jump2014}.
As an exception to the rule, the 1989 exercise was never used in any bibliometric study that compared the peer review results to metrics (although there were other analyses; see, for example, \citet{Johnes1993}).

The 1992 exercise---then called the Research Assessment Exercise (RAE)---sparked more bibliometric interest.
In addition to allowing two publications to be nominated for assessment by the institutions, the exercise also collected information on the total number of publications~\citep{Bence2005}.
No less than seven studies appeared that compared the outcomes of the 1992 RAE to bibliometrics.
\citet{Taylor1994} analysed Business \& Management and found a clear correlation\footnote{
  Various studies have employed a multiple regression framework, and they have typically reported $R^2$ values.
  $R^2$ simply corresponds to the square of the (multiple) correlation.
  To provide unified results, we converted all $R^2$ values to their square root and report $R$ values.
  To be clear, we also provide the originally reported $R^2$ values.
  } 
  based on journal publications ($R^2 \approx 0.8$, $R \approx 0.9$).
\citet{Oppenheim1995} analysed Library \& Information Management, and two years later, \citet{Oppenheim1997} considered Anatomy, Archaeology, and Genetics. 
These two studies used both total citation counts and average citation counts per staff and found clear correlations on the order of $0.7$--$0.8$ for both size-dependent and size-independent metrics and all analysed fields.
Only for Anatomy, the size-independent metric was less clearly aligned with peer review outcomes, with a correlation of $R \approx 0.5$.
\cite{LimBanSeng1995} also analysed Library \& Information Management and found even higher correlations on the order of $0.9$ using both average citations and total citations.
The correlation found by \citet{Colman1995} for Politics was lower, at only $0.5$, where they used the number of publications in high impact journals per staff as a metric.
Finally, \citet{Thomas1998} analysed Business \& Management Studies using a journal-based score and found a correlation of $0.68$.
For the 1992 exercise, overall, both size-dependent and size-independent metrics correlated reasonably well with peer review in quite a number of fields.
Most authors recommended that the RAE should take metrics into account, for example, as an initial suggestion, which can then be revised based on peer review.

In the 1996 RAE, full publication lists were no longer collected~\citep{Bence2005}.
In 2001, results were announced as rankings, and institutions also received an overall score of $1$--$5^*$.
\citet{Smith2002} analysed both the 1996 and the 2001 RAE and found a correlation on the order of $0.9$ for the average number of citations in Psychology for both exercises.
\citet{Clerides2011} also analysed both the 1996 and the 2001 RAE and found a correlation of about $R \approx 0.7$ ($R^2 \approx 0.5$) using the total number of high impact journal articles.
\citet{Norris2003} analysed Archaeology and found correlations of about 0.8 for both size-dependent and size-independent metrics.
\citet{Mahdi2008} analysed all units of assessments (UoAs; i.e. fields) and found that a number of fields showed substantial correlations on the order of $0.7$--$0.8$ (e.g. Clinical Lab. Sciences, Psychology, Biological Sciences, Chemistry, Earth Sciences, and Business \& Management) using the average number of citations per paper.
\citet{Adams2008} also analysed the 2001 RAE results, although their focus was on which granularity of field-normalised citations works best.
They found a reasonably high correlation of about $0.7$ for Psychology, $0.6$ for Physics, and only $0.5$ for Biological Sciences.
Finally, \citet{Butler2009} found a reasonable correlation ($R^2 \approx 0.5 \text{--} 0.6$, $R \approx 0.7 \text{--} 0.8$) for Political Science using the average number of citations.

In 2008, the results of the RAE were more structured.
Rather than providing overall scores for institutions per UoA, a so-called quality profile was provided\footnote{Data on the results and submissions are provided at \href{https://www.rae.ac.uk}{www.rae.ac.uk}}.
The quality profile offered more detailed information on the proportion of outputs that were awarded $1$--$4$ stars.
This enabled a more detailed analysis, since the measure was much more fine grained than the overall outcome.
In addition, it allowed a clear distinction between size-dependent and size-independent results.
Previously, only the overall results were announced, and the extent to which the results were size-dependent or size-independent was unclear.
Most studies found that larger institutions generally did better in RAEs, implying a certain type of size-dependent component, but this was never entirely clear.
From 2008 onwards, the results were announced as a proportion of outputs that were awarded a certain number of stars, which was unambiguously size-independent.

\citet{Norris2010} examined Library \& Information Science, Anthropology, and Pharmacy in the 2008 RAE using the $h$-index (and a variant thereof) and total citation counts.
They compared this to a weighted average of the results multiplied by the number of staff, clearly a size-dependent metric.
\citet{Norris2010} found a correlation of about 0.7 for Pharmacy, while Library \& Information Science showed a correlation of only about 0.4, and Anthropology showed even a negative correlation.
\citet{Taylor2011} analysed Business \& Management, Economics \& Econometrics, and Accounting \& Finance.
They relied on a journal list from UK business schools to determine the proportion of publications in top journals and found a quite high correlation ($R^2 \approx 0.64 \text{--} 0.78$, $R \approx 0.80 \text{--} 0.88$) with the average rating. 
\citet{Kelly2012} found a clear correlation ($R^2 = 0.83$, $R = 0.91$) for Sociology.
They also used the proportion of publications in top journals and compared it to a weighted average of RAE results.
\citet{Mckay2012} found that most scholars in the field of Social Work, Social Policy \& Administration did not necessarily submit their most highly cited work for evaluation.
This study did not explicitly report how well citations match peer review.
\citet{Allen2013} replicated the study of \citet{Butler2009} of Politics \& International Studies and found a similar correlation ($R^2 \approx 0.7$, $R \approx 0.85$).
They correlated the proportion of publications in top journals with the proportion of publications that obtained four stars, which are both clearly size-independent measures.

In two publications, \citet{Mryglod2013,Mryglod2013a} explicitly studied size-dependent correlations versus size-independent correlations in seven fields (Biology, Physics, Chemistry, Engineering, Geography \& Environmental Science, Sociology, and History).
They studied the average normalised citation score and the total normalised citation score and examined how they correlate with the RAE Grade (a weighted average of scores) and the RAE Score (the RAE Grade times the number of staff), respectively.
They found size-independent correlations of only about $0.34$ for History and Engineering and up to about $0.6$ for Biology and Chemistry.
The size-dependent correlations were substantially higher and reached about $0.9$ for all fields.
We discuss this in more detail in Section~\ref{sec:size_dependent}.

In conclusion, most studies in the literature have found correlations on the order of $0.6$--$0.7$ for fields that seem to be amenable to bibliometric analysis.
The conclusions that were drawn from such results nonetheless differed.
Three types of conclusions can be distinguished.
First, some authors concluded simply that the observed correlation was sufficiently high to replace peer review by metrics.
Others concluded that peer review should be supported by citation analysis.
Finally, some concluded that peer review should not be replaced by metrics, even though they found relatively high correlations.
This indicates that different researchers draw different conclusions, despite finding similar correlations.
One problem is that none of the correlations are assessed against the same yardstick; thus, it is unclear when a correlation should be considered ``high'' and when it should be considered ``low''.

\subsection{Research Excellence Framework 2014}

\noindent The REF 2014 was accompanied by an extensive study into the possibilities of using metrics instead of peer review, known as the \emph{Metric Tide} report~\citep{Wilsdon2015}.
This report concluded that citations should only supplement, rather than supplant, peer review.
One of the arguments for this conclusion was based on an analysis of how field-normalised citations based on Scopus data correlate with peer review.
The report found correlations\footnote{The report also used precision and specificity, which are more appropriate than correlations for the individual publication level, but for comparability, we here focus on the reported correlations.} in the range of about $0.2$--$0.4$.
This is quite low compared with most preceding studies, which found correlations of roughly $0.6$--$0.7$.
In contrast to preceding studies, \citet[Supplementary Report II]{Wilsdon2015} analysed the correlation between metrics and peer review at the level of individual publications rather than at some aggregate level.
This is an important difference that we revisit in Section~\ref{sec:level_of_aggregation}.

The REF 2014 results were also analysed by~\citet{Mryglod2015,Mryglod2015a} at the institutional level.
They found that the departmental $h$-index was not sufficiently predictive, even though an earlier analysis suggested that the $h$-index might be predictive in Psychology~\citep{Bishop2014}.
An analysis by Elsevier found that metrics were reasonably predictive of peer review outcomes at an institutional level in some fields but not in others~\citep{Jump2015}.

Both \citet{Pride} and \citet{Harzing2017} compared the UK REF results with metrics using Microsoft Academic Graph~\citep{Harzing2017a}.
\citet{Pride} compared the median number of citations with the REF GPA, which is a weighted average of the proportion of publications that have been awarded a certain number of stars for all UoAs, clearly taking a size-independent perspective.
They found correlations on the order of $0.7$--$0.8$ for the UoAs that showed the highest correlations.
\citet{Harzing2017} compared the total number of citations and a so-called REF power rating, taking a size-dependent perspective, and found a very high correlation of $0.97$.
This correlation was obtained at an even higher aggregate level, namely, the overall institutional level, without differentiating between different disciplines.
She found similarly high correlations when studying Chemistry, Computer Science, and Business \& Management separately.
The high correlations can be partly explained by the use of a size-dependent perspective.
We comment on this in Section~\ref{sec:size_dependent}.

\section{Data and Methods}

\noindent The REF 2014 provides a well-documented dataset of both the evaluation results and the submitted publications that have been evaluated\footnote{All data can be retrieved at \href{http://www.ref.ac.uk/2014/}{www.ref.ac.uk/2014}.}.
The REF 2014 has different scores for different profiles: ``output'', (societal) ``impact'', and ``research environment''.
Only the ``output'' profile is based on an evaluation of the submitted publications.
The others are based on case studies and other (textual) materials.
We restrict ourselves to the REF 2014 scores in the output profile, and we compare them with citation metrics.

We match publications to the CWTS in-house version of the Web of Science (WoS) through their DOI.
We use the Science Citation Index Expanded, the Social Sciences Citation Index, and the Arts \& Humanities Citation Index.
Most publications are articles (type `D' in the REF 2014 dataset), but the publications also include books, conference proceedings, and other materials.
In total, $190\,962$ publications were submitted, of which $149\,616$ have an associated DOI, with $133\,469$ of these being matched to the WoS.
Overall, the WoS covers about two-thirds of all submitted publications.
Some fields are poorly covered in the WoS, such as the arts and humanities, having a coverage of only about $10$--$30\%$ of submitted publications, whereas the natural sciences generally have a high coverage of $90$--$95\%$ (see Table~\ref{tab:estimate_b} for an overview).
In the calculation of citation metrics, we take into account only publications covered in the WoS.
In the calculation of statistics based on peer review, all publications submitted to the REF are considered, including those not covered in the WoS.

All matched publications are associated with a particular UoA, which roughly corresponds to a field or discipline.
The REF 2014 distinguished $36$ UoAs.
Every publication was submitted on behalf of a particular institution.
Some publications were submitted in multiple UoAs, and we take them into account in each UoA.
Publications that were co-authored and submitted by multiple institutions may thus be counted multiple times.
Publications co-authored by several authors from the same institution were sometimes submitted multiple times in the same UoA by the same institution\footnote{Occasionally, incorrect DOIs were provided, resulting in seemingly duplicate publications for the same UoA and institution.}.
We consider only the unique publications of an institution in a UoA.
In other words, we count a publication only once, even if it was submitted multiple times in the same UoA by the same institution.

Some institutions can have separate submission headings in the same UoA to differentiate more fine-grained subjects.
For example, Goldsmiths' College separately submits publications for Music and Theatre \& Performance in the overall UoA of Music, Drama, Dance \& Performing Arts.
The results of such separate submissions are also announced separately, and we therefore also consider them to be separate submissions.

We consider citations coming from publications up to and including 2014, which is realistic if metrics had actually been used during the REF itself.
For this reason, we exclude $365$ publications that were officially published after 2014 (although they may have already been available online).
We use about $4\,000$ micro-level fields constructed algorithmically on the basis of citation data~\citep{Waltman2012,Ruiz-Castillo2015} to perform field normalisation.
Citations are normalised on the basis of publication year and field, relative to all publications covered in the WoS. 

We calculate how many $4^*$ publications correspond to how many top $10\%$ publications per UoA (see Table~\ref{tab:estimate_b}).
This can differ quite substantially from one UoA to another.
For example, Clinical Medicine shows $0.57$ $4^*$ publications per top $10\%$ publication, whereas Mathematical Sciences show $1.18$ $4^*$ publications per top $10\%$ publication.
This suggests that what is considered as $4^*$ publication in peer review differs per field, where some fields seem to use more stringent conditions than others.
Similarly, \citet{Wooding2015} found that peer review was less stringent in REF 2014 than in REF 2008, in what publications were considered worthy of $4^*$.

Before presenting our results, we first address four important methodological considerations.
We start by reflecting on the level of aggregation at which agreement between metrics and peer review should be analysed.
We then examine both the size-dependent and size-independent perspectives, especially regarding correlations.
This leads us to consider alternative measures of agreement.
Finally, we discuss the matter of peer review uncertainty.

\subsection{Level of aggregation}
\label{sec:level_of_aggregation}

\noindent The \emph{Metric Tide} report analysed agreement between metrics and peer review at the level of individual publications.
We believe that this is not appropriate in the context of the REF, and it may explain the large differences between the \emph{Metric Tide} report and preceding publications in which agreement between metrics and peer review was analysed.
The institutional level is the appropriate level to use for the analysis.
The analysis at the level of individual publications is very interesting.
The low agreement at the level of individual publications supports the idea that metrics should generally not replace peer review in the evaluation of a single individual publication.
However, the goal of the REF is not to assess the quality of individual publications but rather to assess ``the quality of research in UK higher education institutions''\footnote{\url{https://www.ref.ac.uk/about/}}.
Therefore, the question should not be whether the evaluation of individual publications by peer review can be replaced by the evaluation of individual publications by metrics but rather whether the evaluation of institutions by peer review can be replaced by the evaluation of institutions by metrics.
Even if citations are not sufficiently accurate at the individual publication level, they could still be sufficiently accurate at the aggregate institutional level; the errors may `cancel out'.
For this reason, we perform our analysis at the institutional level.
We calculate citation metrics per combination of an institution and a UoA.

\subsection{Size-dependent and size-independent perspectives}
\label{sec:size_dependent}

\noindent As briefly discussed earlier, the REF has multiple objectives.
It aims to provide a reputational yardstick, but it also aims to provide a basis for distributing funding.
A reputational yardstick is usually related to the average scientific quality of the publications of an institution in a certain UoA.
As such, a reputational yardstick is \emph{size-independent}: it concerns an average or percentage, not a total, and it does not depend on the size of an institution.
In the REF, funding is related to the total scientific quality of the publications of an institution in a certain UoA.
As such, funding is \emph{size-dependent}: institutions with more output or staff generally receive more funding.
Of course, quality also affects funding: institutions that do well receive more funding than equally sized institutions that do less well.
Both the size-dependent and size-independent perspectives are relevant to the REF.
We therefore believe that both perspectives are important in deciding whether metrics can replace peer review.

Many studies of the REF and its predecessors have analysed either size-dependent or size-independent correlations.
Size-dependent correlations are typically much higher than size-independent correlations. 
For example, \citet{Mryglod2013,Mryglod2013a} found size-dependent correlations on the order of $0.9$ but much lower correlations for size-independent metrics.
Similarly, \citet{Harzing2017} found a very high size-dependent correlation.

\begin{figure}[tb]
  \centering
  \includegraphics[width=\linewidth]{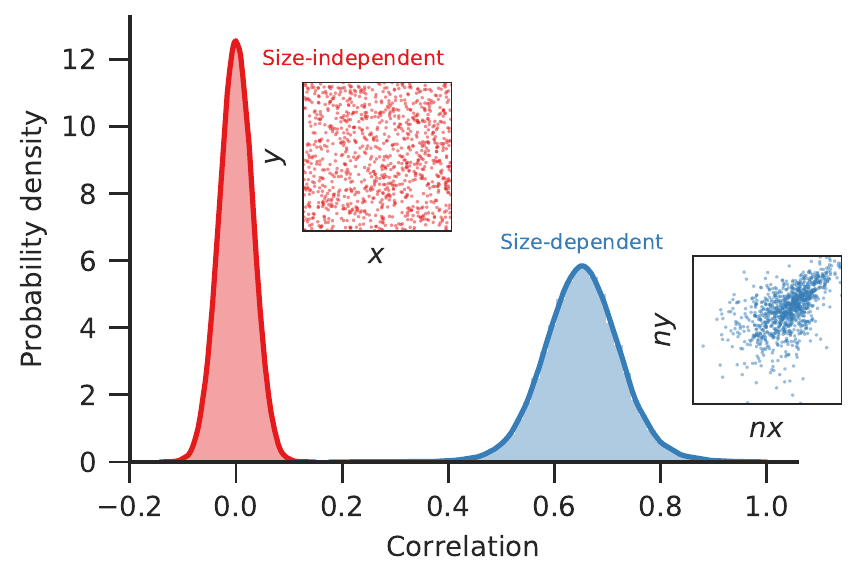}
  \caption{
    The correlation between two size-dependent metrics can be quite high even if the corresponding size-independent metrics are completely uncorrelated.
    The insets show the scatter plots of size-dependent and size-independent metrics.
    For the size-independent scatter plot logarithmic scales are used.
  }
  \label{fig:spurious}
\end{figure}

Size-dependent correlations can be expected to be larger in general.
Let us make this a bit more explicit.
Suppose we have two size-independent metrics $x$ and $y$ (e.g. metrics and peer review), where $n$ denotes the total size (e.g. number of publications or staff).
The two size-dependent metrics would then be $x n$ and $y n$.
Then, even if $x$ and $y$ are completely independent from each other, and hence show a correlation of $0$, the two size-dependent metrics $x n$ and $y n$ may show a quite high correlation.
This is demonstrated in Fig.~\ref{fig:spurious}, where $x$ and $y$ are two independent uniform variables and $n$ is a standard log-normal variable ($1000$ samples).
In this example, the Pearson correlation between $x n$ and $y n$ may be as high as $0.7$--$0.8$.
In other words, the fact that $x n$ and $y n$ may show a high correlation may be completely explained by the common factor $n$.
A similar observation has already been made before in bibliometrics~\citep{West2010}, and related concerns were already raised by~\citeauthor{Pearson1896} as early as 1896.

This complicates the interpretation of size-dependent correlations.
A high size-dependent correlation may be due to $x$ and $y$ being strongly correlated, but it may also be due to $n$ having a high variance.
The higher the variance of $n$, the higher the size-dependent correlation.
In fact, if $n$ is distributed according to a log-normal distribution with a very large variance, the size-dependent correlation will be close to $1$, regardless of the extent to which $x$ and $y$ are correlated.
The strength of the size-dependent correlation then mainly reflects the variance of the size of institutions.

In our analysis, we consider both a size-dependent and a size-independent perspective.
We calculate the proportion of publications that belong to the top $10\%$ most highly cited publications in their field and year, which we call the PP(top $10\%$).
In addition, we use PP($4^*$) to denote the proportion of publications with a $4^*$ rating in the REF.
The PP(top $10\%$) and PP($4^*$) are similar in spirit.\footnote{Note that PP(top $10\%$) concerns the proportion of publications that have been matched in the WoS, whereas PP($4^*$) concerns the proportion of all submitted outputs.}
They aim to identify whether publications have a high impact or are of high quality (``world leading''), respectively.
Other citation metrics, such as those based on average normalised citation counts, are more difficult to translate into a $4^*$ rating system.
Both the PP($4^*$) and the PP(top $10\%$) are clearly size-independent.
We calculate the total number of $4^*$ rated outputs, called the P($4^*$), by multiplying the PP($4^*$) by the number of submitted outputs.
Similarly, we obtain the total number of top $10\%$ outputs, called the P(top $10\%$), by multiplying the PP(top $10\%$) by the number of submitted publications in the WoS.
Both the P($4^*$) and the P(top $10\%$) are clearly size-dependent.

\subsection{Measures of agreement}
\label{sec:measures_main}

\noindent Agreement between metrics and peer review can be measured using a variety of measures.
For example, the \emph{Metric Tide} report employs measures such as precision and sensitivity, which are well suited for the individual publication level.
Most analyses of the REF and its predecessors employ correlation coefficients.
As we argued in the previous section, correlations may be difficult to interpret when taking a size-dependent perspective.
Moreover, correlations provide little intuition of the size of the differences between metrics and peer review. 
For this reason, we consider two different measures (see Appendix~\ref{sec:measures} for details): the median absolute difference (MAD) and the median absolute percentage difference (MAPD).

The MAD gives an indication of the absolute differences that we can expect when switching from peer review to metrics.
We believe that this measure is especially informative when taking a size-independent perspective.
For example, if an institution has a PP($4^*$) of $30\%$ and the MAD is $3$ percentage points, then in half of the cases switching to metrics would yield an outcome equivalent to a PP($4^*$) between $27$ and $33\%$.
The idea of the MAD is that an increase or decrease of $3$ percentage points would likely be of similar interest to institutions with different PP($4^*$) scores.
That is, if one institution has a PP($4^*$) of $50\%$ and another has a PP($4^*$) of $30\%$, a difference of $3$ percentage points would be of similar interest to both.

This is quite different for the size-dependent perspective.
The size of institutions varies much more than the proportion of $4^*$ publications of institutions.
As such, a certain absolute difference will probably not be of the same interest to different institutions when taking a size-dependent perspective.
For example, in terms of funding, if we report an absolute difference of \pounds$10\,000$, this would be of major interest to institutions receiving only \pounds$20\,000$, but probably not so much for institutions receiving \pounds$1\,000\,000$.
From this point of view, the MAPD can be considered more appropriate, as it gives an indication of the relative differences that we can expect when switching from peer review to metrics.
The idea of MAPD is that an increase or decrease of $10\%$ would likely be of similar interest to both small institutions that receive little funding and large institutions that receive much funding.
The MAPD is the same for both size-dependent and size-independent metrics, since the common factor falls out in the calculation (see Appendix~\ref{sec:measures} for details).

\subsection{Peer review uncertainty}

\noindent Regardless of the measure of agreement, the perspective (i.e. size-independent or size-dependent), and the level of aggregation, it is important to acknowledge that peer review is subject to uncertainty.
Hypothetically, if the REF peer review had been carried out twice, based on the same publications but with different experts, the outcomes would not have been identical.
This is what we refer to as peer review uncertainty.
It is sometimes also called internal peer review agreement.
Evidence from the Italian research assessment exercise, known as the VQR, suggests that peer review uncertainty is quite high~\citep{Bertocchi2015}.
Unfortunately, detailed peer review results of the REF at the publication level are not available.
Also, the \emph{Metric Tide} report~\citep{Wilsdon2015} did not quantify internal peer review agreement, which could have served as a baseline for our study.
Internal peer review agreement in the REF has not been investigated in other publications either, although peer review in the REF has been studied from other perspectives~(e.g. \citealt{Derrick2018}).

To quantify peer review uncertainty and get an idea of the order of magnitude of the agreement that we can expect in peer review itself, we perform a type of bootstrap analysis (see Appendix~\ref{sec:model} for details).
Since we do not know exactly the degree of uncertainty in peer review, we consider two scenarios, one with low uncertainty ($\sigma_\epsilon^2 = 0.1$, see Appendix~\ref{sec:model}) and one with high uncertainty ($\sigma_\epsilon^2 = 1$).
The results presented in the next section are based on $1\,000$ bootstrap samples.
We report both the median outcome obtained from $1\,000$ samples and the interval that covers $95\%$ of the outcomes.

\section{Results}
\label{sec:results}

\noindent We now describe the results from our analysis.
Our analysis compares the agreement between metrics and peer review with the internal agreement of peer review, based on a simple model of peer review. 
For simplicity, we consider only $4^*$ publications, as they are deemed four times more valuable than $3^*$ publications in the REF.
We first describe our results from the size-independent perspective and then turn to the size-dependent perspective.
All necessary replication materials have been deposited at Zenodo~\citep{replication} and can be accessed at \url{https://github.com/vtraag/replication-uk-ref-2014}.

\subsection{Size-independent perspective}

\begin{figure}[t]
  \centering
  \includegraphics[width=\linewidth]{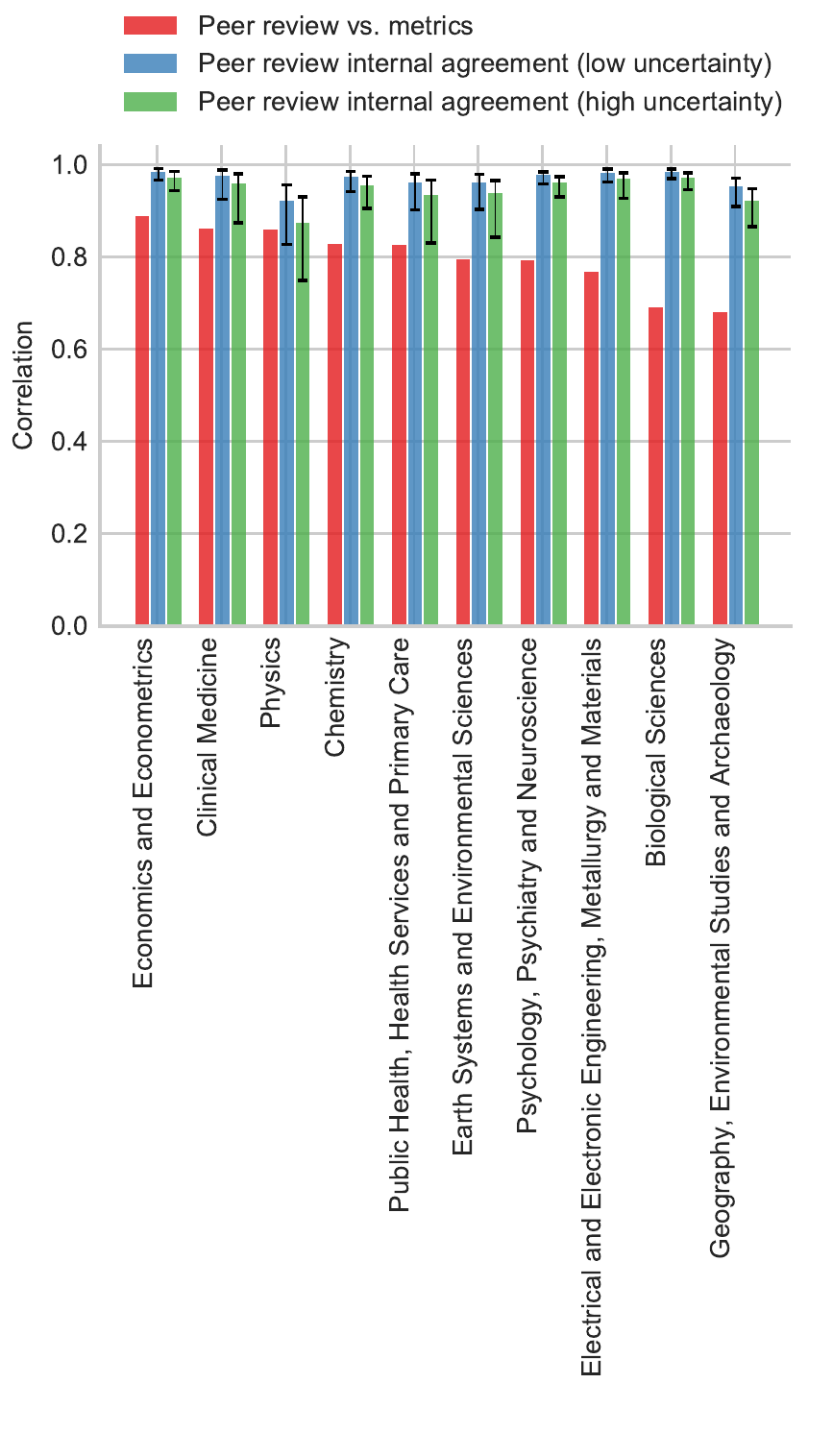}
  \caption{Size-independent correlation between $\PP(\text{top}~10\%)$ and $\PP(4^*)$ compared with correlations based on a model of peer review uncertainty.
  Results are shown only for the $10$ units of assessment with the highest correlation between metrics and peer review.}
  \label{fig:correlation_barchart_size_indep}
\end{figure}

\begin{figure}[t]
  \centering
  \includegraphics[width=\linewidth]{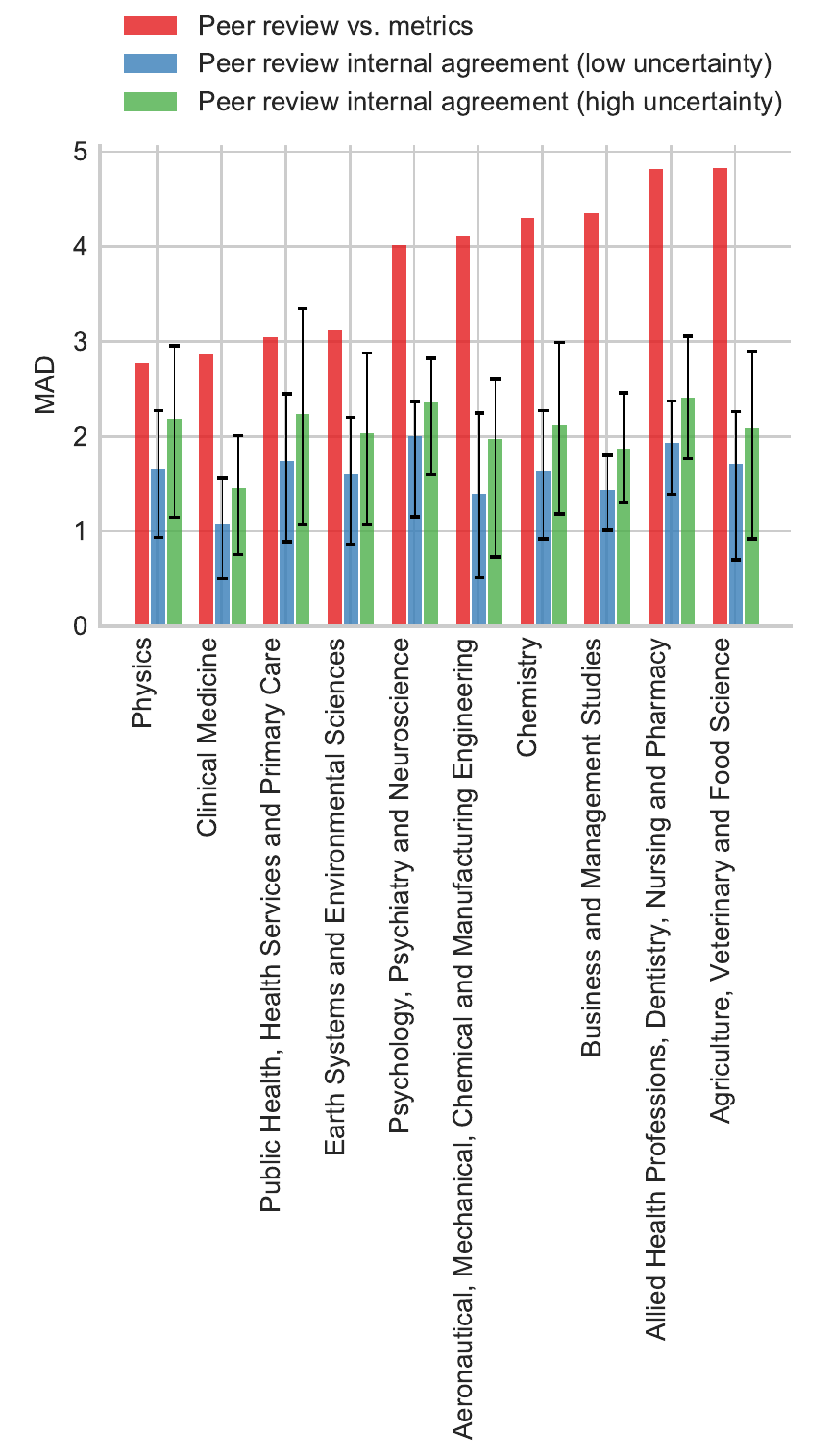}
  \caption{Size-independent median absolute difference (MAD) between $\PP(\text{top}~10\%)$ and $\PP(4^*)$ compared with the MAD based on a model of peer review uncertainty.
  Results are shown only for the $10$ units of assessment with the lowest MAD between metrics and peer review.}
  \label{fig:MAD_barchart_size_indep}
\end{figure}

\begin{figure}[t]
  \centering
  \includegraphics{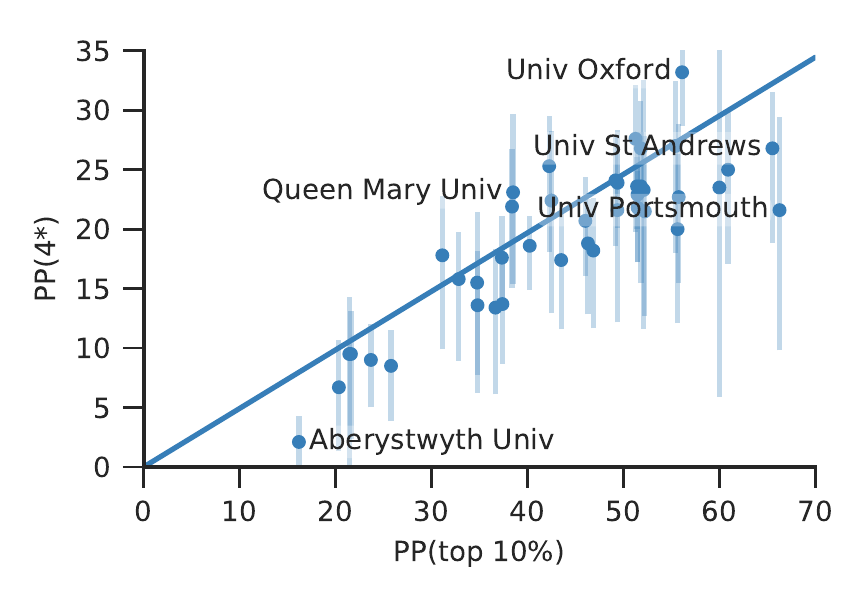}
  \caption{Scatter plot of $\PP(\text{top}~10\%)$ and $\PP(4^*)$ at the institutional level for Physics. 
  Error bars indicate the $95\%$ interval of bootstrapped peer review results for low peer review uncertainty.
  The solid line indicates the proportion of $4^*$ publications considered to be equivalent to a given proportion of top $10\%$ publications (see Appendix~\ref{sec:measures} for details).
  }
  \label{fig:scatter_size_indep_physics}
\end{figure}

\noindent To facilitate comparison with earlier studies, we first discuss our results in terms of Pearson correlations.
We find that Economics \& Econometrics, Clinical Medicine, Physics, Chemistry, and Public Health show a high size-independent Pearson correlation between the percentage of $4^*$ rated submissions and the percentage of top $10\%$ publications: Pearson correlations are higher than $0.8$ (see Fig.~\ref{fig:correlation_barchart_size_indep} and Table~\ref{tab:agreement}).
A number of other fields show correlations on the order of $0.7$, which is in line with previous studies on earlier rounds of the RAE/REF.
These correlations are much higher than the correlations found by the \emph{Metric Tide} report~\citep{Wilsdon2015}.

Our results strongly differ from the analysis by Elsevier of the REF results~\citep{Jump2015}, even though it also found some relatively strong correlations.
In particular, the analysis found correlations for Physics and Clinical Medicine on the order of $0.3$.
Public Health did a little better, but still the correlation was only about $0.5$.
Finally, Biology had the single highest correlation of about $0.75$, whereas this correlation is much lower in our results.
It may be of interest to compare the different results in more detail and to better understand why Elsevier's results~\citep{Jump2015} differ from ours.
The differences most likely stem from the use of all publications of an institution versus only the publications submitted to the REF.
Another reason for the differences may be the use of different databases (Scopus vs. WoS) and the use of different field classification systems in the field-normalised citation metrics.
The citation metrics of~\citet{Jump2015} were normalised on the basis of the journal-based classification system of Scopus, whereas we normalised on the basis of a detailed publication-based classification system~\citep{Ruiz-Castillo2015}.

The results of the peer review uncertainty may be surprising (see Fig.~\ref{fig:correlation_barchart_size_indep}).
Although the bootstrapped correlations are almost always higher than the correlations of the REF results with the PP(top $10\%$), the differences are sometimes small.
Most notably, Physics shows a correlation between metrics and peer review of $0.86$, which is on par with the bootstrapped correlations, especially for high peer review uncertainty.
This indicates that for Physics, metrics work at least equally well as peer review, assuming some uncertainty in peer review.
For Economics, Clinical Medicine, Chemistry, and Public Health, the correlations between metrics and peer review are lower than the bootstrapped correlations, but the differences are not very large.
Hence, the metrics correlate quite well with peer review for these fields.
Other UoAs show correlations between metrics and peer review that are substantially lower than the correlations obtained using the bootstrapping procedure.

The MAD provides a more intuitive picture of what these correlations mean in practice (see Fig.~\ref{fig:MAD_barchart_size_indep} and Table~\ref{tab:agreement}).
In the interpretation of the MAD, it is important to keep in mind that overall about $30\%$ of the publications have been awarded $4^*$ in the REF.
The MAD in Physics reaches almost $3$ percentage points in PP($4^*$) when switching from peer review to metrics.
This is just somewhat more than $1$ percentage point higher than the median bootstrapped MAD for low peer review uncertainty and less than $1$ percentage point higher than the median bootstrapped MAD for high peer review uncertainty.
Hence, in Physics, the difference between metrics and peer review seems to be just slightly larger than the difference between different peer review exercises.
Moreover, for high peer review uncertainty, the difference between metrics and peer review still falls within the $95\%$ interval of bootstrapped peer review results.
This means that it is possible that the difference between metrics and peer review is of a similar magnitude as the difference between different peer review exercises.
In Clinical Medicine, we also find an MAD of almost $3$ percentage points in PP($4^*$) when switching to metrics, although in this UoA the difference with the bootstrapped MADs is more substantial.
In Public Health, the MAD is slightly higher than $3$ percentage points in PP($4^*$).
The difference with the bootstrapped MADs is not very large, and for high peer review uncertainty, it falls within the $95\%$ interval of bootstrapped peer review results.
For other fields, we observe that the MAD when switching from peer review to metrics is higher than the bootstrapped MADs, but for many of these fields, the MAD may still be considered to be relatively small (e.g. $< 5$ percentage points).
On the other hand, there are also fields for which the MAD is quite large (see Fig.~\ref{fig:MAD_barchart_size_indep_all} for the MADs for all UoAs).
These are especially fields that are not well covered in the WoS.

Looking at the result for Physics in more detail, we see that most institutions have bootstrapped peer review results that agree reasonably well with metrics (see Fig.~\ref{fig:scatter_size_indep_physics}, see Fig.~\ref{fig:scatter_size_indep} for all UoAs).
However, some larger differences remain.
University of Oxford and Queen Mary University are systematically valued more highly by peer review than by metrics. 
Conversely, University of St Andrews, University of Portsmouth, and Aberystwyth University are systematically valued less highly by peer review than by metrics.

\subsection{Size-dependent perspective}

\begin{figure}[t]
  \centering
  \includegraphics[width=\linewidth]{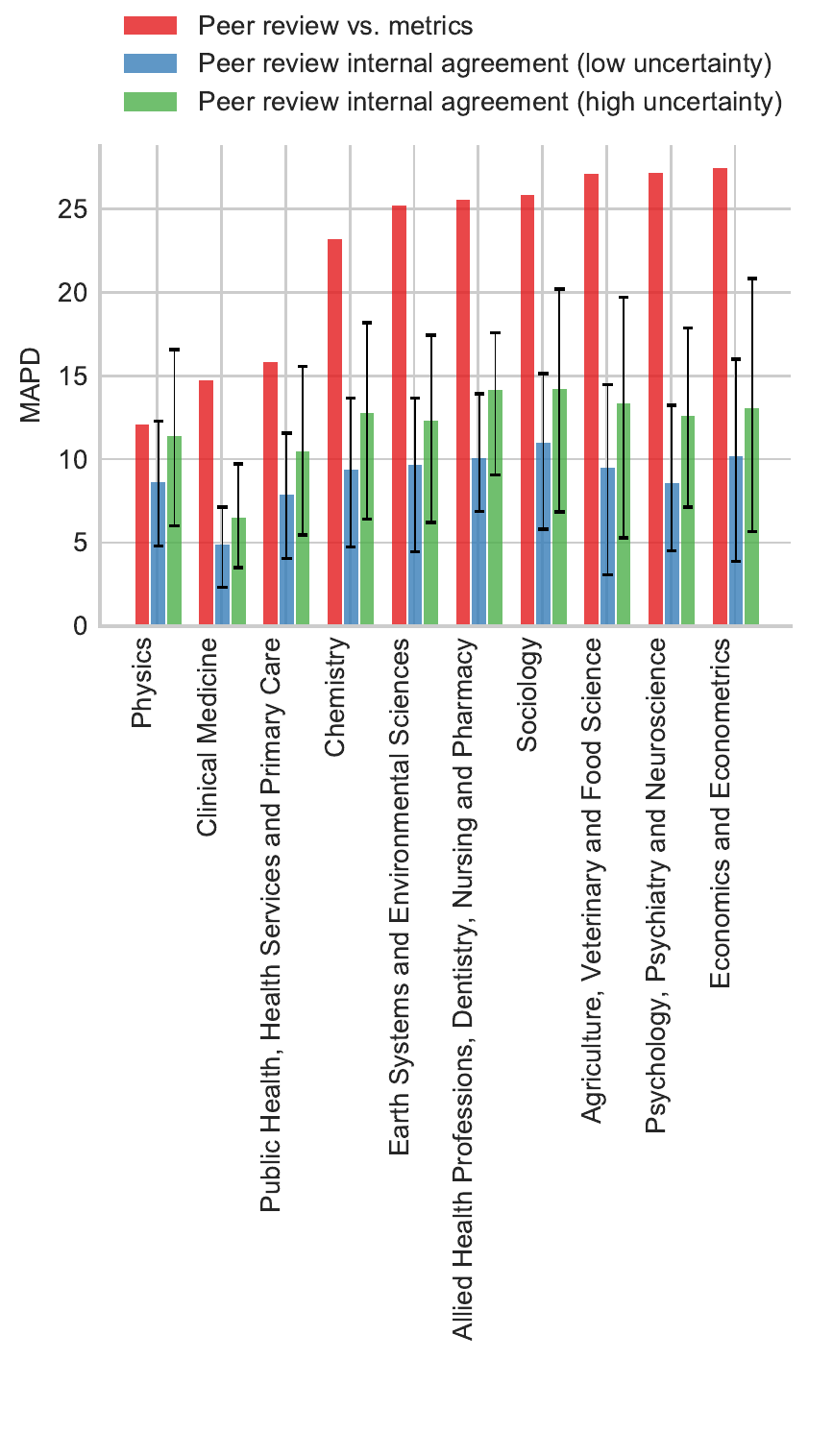}
  \caption{Size-dependent median absolute percentage difference (MAPD) of $\Pub(\text{top}~10\%)$ relative to $\Pub(4^*)$ compared with the MAPD based on a model of peer review uncertainty.
  Results are shown only for the $10$ units of assessment with the lowest MAPD of metrics relative to peer review.}
  \label{fig:MAPD_barchart_size_dep}
\end{figure}

\noindent As expected, the size-dependent correlations are much higher than the size-independent correlations (see Table~\ref{tab:agreement}).
Half of all UoAs reach correlations higher than $0.9$.
Some have very high size-dependent correlations, even when the size-independent correlations are low, as previously explained in Section~\ref{sec:size_dependent}.
For example, Mathematical Sciences shows a size-dependent correlation of $0.96$, whereas the size-independent correlation is only $0.39$.
As discussed above, we believe the correlations are not so informative for the size-dependent perspective, and we therefore focus on the MAPD.

Peer review uncertainty leads to MAPDs of somewhere between $10\%$ and $15\%$ for many fields (see Fig.~\ref{fig:MAPD_barchart_size_dep}).
Hence, peer review uncertainty may have a substantial effect on the amount of funding allocated to institutions.
Comparing peer review with metrics, we find that Physics has an MAPD of $12\%$, which is similar to what can be expected from peer review uncertainty.
Clinical Medicine has an MAPD of almost $15\%$, which is substantially higher than the MAPD resulting from peer review uncertainty.
Likewise, Public Health has an MAPD of about $16\%$, which is higher than the expectation from peer review uncertainty.
Other fields show MAPDs between metrics and peer review that are above $20\%$, especially fields that are not well covered in the WoS (see Fig.~\ref{fig:MAPD_barchart_size_dep_all}).
The $10$ UoAs with the lowest MAPDs all show size-dependent correlations close to or above $0.9$, which illustrates how correlations and MAPDs may potentially lead to different conclusions.
Biology is a clear example: it has a size-dependent correlation of $0.98$, yet it has an MAPD of $32\%$.

\begin{figure}[t]
  \centering
  \includegraphics[width=\linewidth]{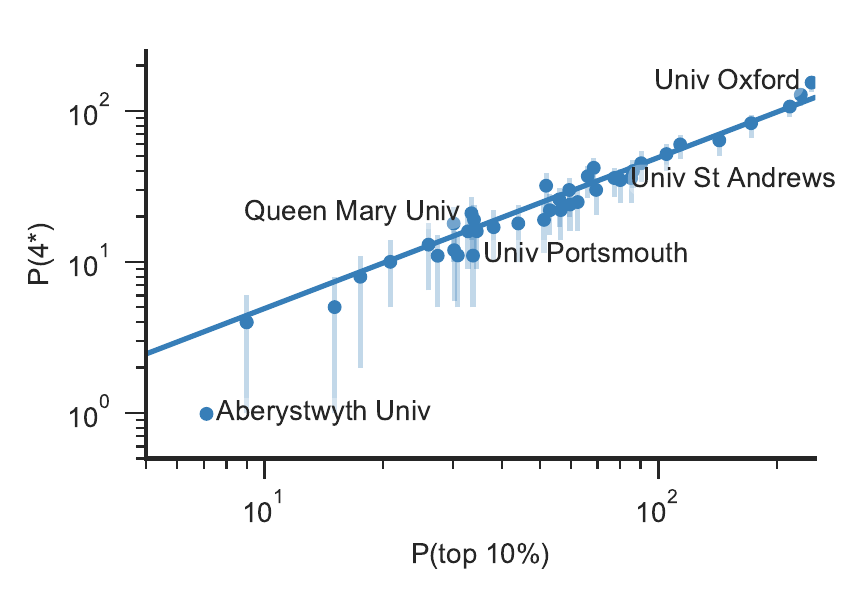}
  \caption{Logarithmic scatter plot of $\Pub(\text{top}~10\%)$ and $\Pub(4^*)$ at the institutional level for Physics. 
  Error bars indicate the $95\%$ interval of bootstrapped peer review results for low peer review uncertainty.
  The solid line indicates the number of $4^*$ publications considered to be equivalent to a given number of top $10\%$ publications (see Appendix~\ref{sec:measures} for details).
  }
  \label{fig:scatter_size_dep_physics}
\end{figure}

The MAPD summarises the overall differences, but for individual institutions, the differences can be substantially larger or smaller.
We again consider Physics in somewhat more detail (see Fig.~\ref{fig:scatter_size_dep_physics}, see Fig.~\ref{fig:scatter_size_dep} for all UoAs).
Although the absolute differences are sometimes difficult to discern in Fig.~\ref{fig:scatter_size_dep_physics}, some of the institutions that we already encountered when taking the size-independent perspective (see Fig.~\ref{fig:scatter_size_indep_physics}) still show clear differences.
University of Oxford would have $22\%$ fewer $4^*$ publications based on metrics than based on peer review, while the difference varies between $-14\%$ and $+9\%$ based on low peer review uncertainty.
Likewise, Queen Mary University would have $22\%$ fewer $4^*$ publications based on metrics than based on peer review.
Based on low peer review uncertainty, the difference varies between $-33\%$ and $+28\%$.
The University of Portsmouth would have $51\%$ more $4^*$ publications based on metrics than based on peer review, while the difference varies between $-55\%$ and $+36\%$ based on low peer review uncertainty.
The University of St Andrews would have $13\%$ more $4^*$ publications based on metrics.
This is within the range of $-30\%$ to $+18\%$ obtained based on low peer review uncertainty.
Finally, Aberystwyth University would have $255\%$ more $4^*$ publications based on metrics, and it would have about $\pm 100\%$ $4^*$ publications based on low peer review uncertainty.
Although other institutions also show differences between metrics and peer review, these are not much larger or smaller than what could be expected based on peer review uncertainty.

\section{Discussion}

\noindent National research assessment exercises evaluate the scientific performance of universities and other research institutions.
To a large extent, this is often based on scientific publications.
The role of citation metrics is regularly discussed in the literature, and the extent to which they correlate with peer review has been repeatedly analysed. 
Recently, in the context of the Research Excellence Framework (REF) 2014 in the UK, the influential \emph{Metric Tide} report~\citep{Wilsdon2015} concluded that metrics should only supplement, rather than supplant, peer review.
The report's conclusion was substantiated by its finding that metrics correlate poorly with peer review.
In contrast, earlier studies have shown that metrics may correlate quite well with peer review. 

The discussion on metrics and peer review is characterised by a variety of correlations and an even larger variety of interpretations of these correlations.
Correlations between metrics and peer review in the \emph{Metric Tide} report are generally on the order of $0.4$.
Most previous studies have found correlations on the order of $0.7$, but some have even reported correlations up to $0.9$.
Conclusions vary, even if the correlations are the same: some argue that a correlation of $0.7$ is too low to consider replacing peer review by metrics, whereas others argue that a correlation of $0.7$ is sufficiently high to do so.

We identify four points that need careful consideration in discussions on the agreement between metrics and peer review: (1) the level of aggregation; (2) whether a size-dependent perspective or a size-independent perspective is taken; (3) appropriate measures of agreement; and (4) uncertainty in peer review.

Most previous studies have analysed the agreement between metrics and peer review at the institutional level, whereas the recent \emph{Metric Tide} report analysed the agreement at the level of individual publications.
For the purpose of deciding between the use of metrics or peer review in the REF, the value of such a publication-level analysis is limited.
The REF results are made available at the institutional level, which is therefore the most appropriate level of analysis.
If correlations at the publication level are low, this does not necessarily mean that correlations at the institutional level will be low as well.
Indeed, we find correlations at the institutional level that are substantially higher than the correlations at the publication level reported in the \emph{Metric Tide} report.
In line with previous results, we obtain size-independent correlations above $0.8$ for a number of fields.

The REF has multiple objectives.
It aims to provide a reputational yardstick, which is, for example, visible in the various league tables that are produced on the basis of the REF.
It also aims to provide a basis for distributing funding.
The objective of a reputational yardstick corresponds to a size-independent perspective, while the objective of funding allocation corresponds to a size-dependent perspective.
Both perspectives are important in deciding whether metrics can replace peer review.

Some authors have found high size-dependent correlations, on the order of $0.9$.
We indeed find similar size-dependent correlations for many fields.
It is important to realise that size-dependent correlations tend to reach high levels because metrics and peer review share a common factor, namely the size of an institution.
This explains why size-dependent correlations may be as high as $0.9$ while the corresponding size-independent correlations may be much lower.
For example, we find a size-dependent correlation of $0.96$ for Mathematical Sciences, whereas the size-independent correlation is only $0.39$.

Measures of agreement should quantify agreement in a way that is most relevant in the specific context in which the measures are used.
From this point of view, correlations are not necessarily the most appropriate measure of agreement.
To compare metrics and peer review, we therefore use two other measures of agreement: the median absolute difference for the size-independent perspective and the median absolute percentage difference for the size-dependent perspective.
In the REF, about $30\%$ of the publications have been awarded $4^*$.
From the size-independent perspective, we find that a number of fields in the REF show a median absolute difference of about $3$ percentage points between metrics and peer review.
In these fields, when switching from peer review to metrics, the percentage of $4^*$ publications of an institution will typically increase or decrease by about $3$ percentage points.
The median absolute percentage difference between metrics and peer review from the size-dependent perspective is about $15\%$ for these fields.
This essentially means that the amount of funding allocated to an institution will typically increase or decrease by about $15\%$.

Differences between metrics and peer review can be interpreted in various ways.
In this paper, we take peer review as the ``gold standard'' that should be matched as closely as possible by metrics.
In the context of the REF this seems the most relevant perspective, because the REF currently relies on peer review and because the use of peer review in the REF seems to be widely accepted.
However, it is also possible that differences between metrics and peer review indicate that metrics better reflect the ``true'' scientific quality of publications than peer review.
Without an independent third measure that can serve as the ``gold standard'', there is no way of establishing whether metrics or peer review offer a better reflection of scientific quality.

Regardless of the level of aggregation at which agreement between metrics and peer review is analysed and regardless of whether a size-dependent or a size-independent perspective is taken, agreement between metrics and peer review should be placed in an appropriate context.
To determine whether agreement between metrics and peer review should be regarded as high or low, it is essential to make a comparison with internal peer review agreement.
Unfortunately, there are currently no data available to quantify peer review uncertainty in the REF.
Ideally, one needs to have an independent replication of the peer review process in the REF to determine the degree to which peer review is subject to uncertainty and to quantify internal peer review agreement.
We recommend that uncertainty in peer review is analysed in the next round of the REF in 2021 to clarify this important point.

Given the lack of empirical data, we rely on a simple model to get an idea of the degree of uncertainty in peer review.
For some fields, our model suggests that agreement between metrics and peer review is quite close to internal peer review agreement.
In particular, this is the case for Physics, Clinical Medicine, and Public Health, Health Services \& Primary Care.
For these fields, the differences between metrics and peer review are relatively minor, from both a reputational (size-independent) and a funding (size-dependent) perspective.
From the viewpoint of agreement between metrics and peer review, in these fields one may consider switching from peer review to metrics.

In some fields, metrics were used to inform the REF peer review. 
Even in fields in which metrics were not used in a formal way, reviewers may still have informally been influenced by metrics. 
It could be argued that this explains the high agreement between metrics and peer review. 
This may suggest that peer review should be organised differently. 
For example, peer reviewers should have sufficient time to properly evaluate each publication without the need to rely on metrics. 
Still, it may be difficult to limit the influence of metrics. 
Peer reviewers may have a strong tendency to echo what metrics tell them. 
The added benefit of peer review then seems questionable, especially considering the time and money it requires.

Importantly, we do not suggest that metrics \emph{should} replace peer review in the REF.
As shown in this paper, the argument that metrics should not be used because of their low agreement with peer review does not stand up to closer scrutiny for at least some fields.
However, other arguments against the use of metrics may be provided, even for fields in which metrics and peer review agree strongly.
Foremost, by relying on a metric, the goal of fostering ``high quality'' science may become displaced by the goal of obtaining a high metric.
Metrics may invite gaming of citations and strategic behaviour that has unintended and undesirable consequences~\citep{Rijcke2016}.
For example, evaluation on the basis of certain metrics may unjustly favour problematic research methods, which may lead to the ``evolution of bad science''~\citep{Smaldino2016}.
The use of a metric-driven approach in some fields, while maintaining a peer review approach in other fields, may also complicate the evaluation exercise and amplify disciplinary differences.
Other arguments against replacing peer review by metrics are of a more pragmatic or more practical nature.
One argument is that citation analysis may wield insufficient support and confidence in the scientific community~\citep{Wilsdon2015}.
Another argument is that there will always be some outputs that are not covered in bibliographic databases and for which it is not possible to obtain metrics.
Of course, there are also other arguments in favour of metrics.
For example, the total costs of the recent REF 2014 have been estimated at £246 million~\citep{Farla2015}.
By relying on metrics instead of peer review these costs could be reduced.
First of all, the costs of panelists' time (£19 million) could be saved.
However, the bulk of the costs (£212 million) were born by the institutions themselves in preparing the submissions to the REF.
To reduce these costs, it has been suggested to simply consider all publications of institutions rather than only a selection~\citep{Harzing2017}.
All above arguments for and against metrics and peer review should be carefully weighed in the discussion on whether metrics should (partly) replace peer review in the REF.

Finally, as a limitation of our work, we emphasise that we do not consider the broader societal, cultural, and economic impact that is also evaluated in the REF.
Such a broader evaluation cannot be done on the basis of metrics~\citep{Ravenscroft2017,Bornmann2018,Pollitt2016} and should therefore be carried out using peer review.
Outputs that are not covered in bibliographic databases such as the WoS, Scopus, Dimensions, and Microsoft Academic also need to be assessed by peer review.

\begin{acknowledgments}
We thank Lutz Bornmann, Anne-Wil Harzing, Steven Hill, Sven Hug, and David Pride for their comments on an earlier version of this paper.
We like to thank Jeroen Baas for discussion on the analysis by Elsevier.
\end{acknowledgments}

\bibliography{bibliography}

\appendix
\onecolumngrid

\counterwithin{figure}{section}
\counterwithin{table}{section}

\section{Measures of agreement}
\label{sec:measures}

\noindent The Pearson correlation coefficient quantifies the extent to which a linear relationship of the form $\hat{y}_i = a + b x_i$ provides a good fit to the data.
In this framework, both the intercept $a$ and the slope $b$ are estimated based on the least squares principle.
According to this principle, we find $a$ and $b$ such that $\sum_{i} (\hat{y}_i - y_i)^2$ is minimal.
The explained variance $R^2$ can then be expressed as
\begin{equation}
  R^2 = 1 - \frac{\sum_i (\hat{y}_i - y_i)^2}{\sum_i (\bar{y} - y_i)^2},
\end{equation}
where $\bar{y} = \frac{1}{n} \sum_i y_i$ is the average of $y_1, \ldots, y_n$, with $n$ the number of observations.
The Pearson correlation coefficient is either the positive or the negative square root of $R^2$.
In our context, correlations are usually positive (i.e. $b > 0$), which means that the Pearson correlation coefficient is the positive square root of $R^2$.
When the Pearson correlation coefficient is high, the average of the squared differences, i.e. $\frac{1}{n} \sum_i (\hat{y}_i - y_i)^2$, is small relative to the variance, i.e. $\frac{1}{n} \sum_i (\bar{y} - y_i)^2$.
However, if the variance is very large, a high correlation coefficient may be obtained even though the squared differences are still substantial.
This illustrates the underlying problem of size-dependent correlations, as discussed in Section~\ref{sec:size_dependent}.

Another problem of the Pearson correlation coefficient is that it allows for a non-zero intercept $a$.
In our context, having no top $10\%$ publications should correspond to having no $4^*$ publications.
This means that the intercept $a$ should always be zero.
We then work with the simple linear relationship $\hat{y}_i = b x_i$, and we need to estimate only the slope $b$.
The Pearson correlation has the drawback that it allows for a non-zero intercept $a$, for which there is no proper conceptual justification in our context.

As stated in the main text, we use two measures of agreement in addition to correlations: the median absolute difference (MAD) and the median absolute percentage difference (MAPD).
The MAD is defined as 
\begin{equation}
  \MAD = \median_i{| \hat{y}_i - y_i |},
\end{equation}
and the MAPD is defined as 
\begin{equation}
  \MAPD = \median_i{\frac{| \hat{y}_i - y_i |}{y_i}},
\end{equation}
where $\hat{y}_i = b x_i$.
If $\hat{y}_i = y_i = 0$, we define the MAPD to be $0$.
The MAPD is independent of a multiplicative factor: if we multiply each $y_i$ by a certain $n_i$, this does not affect the MAPD.
In other words,
\begin{equation}
  \median_i{\frac{|n_i \hat{y}_i - n_i y_i|}{n_i y_i}} = \median_i{\frac{|\hat{y}_i - y_i|}{y_i}},
\end{equation}
provided that the estimate $\hat{y}_i$ remains unchanged (which it indeed does, given our estimation of $b$, as we discuss next).

We rely on a simple estimation of $b$ that has a straightforward interpretation.
We determine how many $4^*$ publications are worth a single top $10\%$ publication.
To do so, we calculate for each UoA the ratio of the total number of $4^*$ publications and the total number of top $10\%$ publications.
This ratio then provides our estimate of $b$, and it provides a straightforward way to transform a certain number of top $10\%$ publications into a corresponding number of $4^*$ publications.

The estimates of $b$ for each UoA are reported in Table~\ref{tab:estimate_b}.
The number of $4^*$ publications per top $10\%$ publication varies quite substantially over fields.
In some fields, such as Clinical Medicine and Physics, each top $10\%$ publication is worth about $0.5$ $4^*$ publications.
In other fields, such as Economics \& Econometrics, each top $10\%$ publication is worth about $1.5$ $4^*$ publications.
There are also fields, such as Law and Classics, in which the number of $4^*$ publications per top $10\%$ publication is very high, even above $10$.
To some extent, this is caused by the fact that the WoS coverage in these fields is low.
In addition, the criteria for awarding $4^*$ may not be the same across all UoAs, at least not compared with metrics.

\NewDocumentCommand{\rot}{O{90} O{1em} m}{\makebox[#2][l]{\rotatebox{#1}{#3}}}%

{
\rowcolors{2}{white}{Set1-blue!20}
\setlength{\tabcolsep}{8pt}
\LTcapwidth=\textwidth

\begin{longtable}{lp{6cm}rrrrrr}

\caption{Publication numbers per unit of assessment.}\label{tab:estimate_b}\\
\toprule

\rowcolor{white}
{} &                                                            Unit of assessement & \rot{Nb. submissions} & \rot{Avg. Nb. submissions} & \rot{Nb. $4^*$ pub.} & \rot{Nb. top $10\%$ pub.} & \rot{Ratio $\frac{4^*}{\text{top $10\%$}}$} & \rot{WoS coverage} \\
\midrule
\endfirsthead

\toprule

\rowcolor{white}
{} &                                                            Unit of assessement & \rot{Nb. submissions} & \rot{Avg. Nb. submissions} & \rot{Nb. $4^*$ pub.} & \rot{Nb. top $10\%$ pub.} & \rot{Ratio $\frac{4^*}{\text{top $10\%$}}$} & \rot{WoS coverage} \\
\midrule
\endhead
\midrule
\rowcolor{white}
\multicolumn{8}{r}{{Continued on next page}} \\
\midrule
\endfoot

\bottomrule
\endlastfoot
1  & Clinical Medicine                                                             & 13\,400 & 432 & 3\,107 & 5\,385 & 0.58  & 94.8 \\
2  & Public Health, Health Services and Primary Care                               & 4\,881  & 153 & 1\,093 & 2\,020 & 0.54  & 90.0 \\
3  & Allied Health Professions, Dentistry, Nursing and Pharmacy                    & 10\,358 & 111 & 2\,185 & 2\,009 & 1.09  & 91.3 \\
4  & Psychology, Psychiatry and Neuroscience                                       & 9\,126  & 113 & 2\,361 & 2\,858 & 0.83  & 93.2 \\
5  & Biological Sciences                                                           & 8\,608  & 196 & 2\,511 & 3\,137 & 0.80  & 97.3 \\
6  & Agriculture, Veterinary and Food Science                                      & 3\,919  & 135 & 708    & 1\,001 & 0.71  & 95.5 \\
7  & Earth Systems and Environmental Sciences                                      & 5\,249  & 117 & 951    & 2\,037 & 0.47  & 94.8 \\
8  & Chemistry                                                                     & 4\,698  & 127 & 1\,026 & 1\,646 & 0.62  & 98.9 \\
9  & Physics                                                                       & 6\,446  & 157 & 1\,363 & 2\,769 & 0.49  & 95.0 \\
10 & Mathematical Sciences                                                         & 6\,994  & 132 & 1\,562 & 1\,301 & 1.20  & 84.7 \\
11 & Computer Science and Informatics                                              & 7\,651  & 86  & 1\,693 & 842    & 2.01  & 61.1 \\
12 & Aeronautical, Mechanical, Chemical and Manufacturing Engineering              & 4\,143  & 166 & 752    & 707    & 1.06  & 93.3 \\
13 & Electrical and Electronic Engineering, Metallurgy and Materials               & 4\,025  & 109 & 800    & 912    & 0.88  & 92.2 \\
14 & Civil and Construction Engineering                                            & 1\,384  & 99  & 246    & 258    & 0.95  & 89.5 \\
15 & General Engineering                                                           & 8\,679  & 140 & 1\,486 & 1\,624 & 0.91  & 91.6 \\
16 & Architecture, Built Environment and Planning                                  & 3\,781  & 86  & 840    & 299    & 2.81  & 50.0 \\
17 & Geography, Environmental Studies and Archaeology                              & 6\,017  & 81  & 1\,326 & 1\,444 & 0.92  & 72.9 \\
18 & Economics and Econometrics                                                    & 2\,600  & 93  & 715    & 429    & 1.67  & 78.3 \\
19 & Business and Management Studies                                               & 12\,202 & 125 & 2\,500 & 1\,692 & 1.48  & 77.7 \\
20 & Law                                                                           & 5\,522  & 86  & 1\,104 & 112    & 9.84  & 18.0 \\
21 & Politics and International Studies                                            & 4\,365  & 79  & 910    & 436    & 2.08  & 50.4 \\
22 & Social Work and Social Policy                                                 & 4\,784  & 77  & 917    & 346    & 2.65  & 54.8 \\
23 & Sociology                                                                     & 2\,630  & 91  & 514    & 280    & 1.84  & 56.5 \\
24 & Anthropology and Development Studies                                          & 2\,013  & 81  & 385    & 244    & 1.58  & 49.0 \\
25 & Education                                                                     & 5\,519  & 73  & 1\,205 & 420    & 2.87  & 48.2 \\
26 & Sport and Exercise Sciences, Leisure and Tourism                              & 2\,757  & 55  & 541    & 445    & 1.22  & 76.7 \\
27 & Area Studies                                                                  & 1\,724  & 75  & 408    & 77     & 5.30  & 32.3 \\
28 & Modern Languages and Linguistics                                              & 4\,932  & 95  & 1\,204 & 149    & 8.09  & 23.6 \\
29 & English Language and Literature                                               & 6\,923  & 82  & 1\,950 & 153    & 12.76 & 15.3 \\
30 & History                                                                       & 6\,431  & 79  & 1\,765 & 334    & 5.28  & 26.1 \\
31 & Classics                                                                      & 1\,386  & 63  & 410    & 32     & 12.62 & 8.8 \\
32 & Philosophy                                                                    & 2\,173  & 54  & 569    & 184    & 3.09  & 41.5 \\
33 & Theology and Religious Studies                                                & 1\,558  & 54  & 356    & 18     & 19.76 & 14.1 \\
34 & Art and Design: History, Practice and Theory                                  & 6\,321  & 87  & 1\,130 & 83     & 13.58 & 12.5 \\
35 & Music, Drama, Dance and Performing Arts                                       & 4\,246  & 52  & 1\,071 & 82     & 13.07 & 14.6 \\
36 & Communication, Cultural and Media Studies, Library and Information Management & 3\,517  & 54  & 818    & 140    & 5.86  & 27.0 \\
\end{longtable}

}

\section{Model of REF peer review}
\label{sec:model}

\noindent To analyse uncertainty in peer review in the REF, we use a simple mathematical model. Below, we discuss the details of this model and the way in which the model is used to analyse uncertainty in REF peer review.

\subsection{Model}

\noindent Each institution $k$ submits a certain number of publications $n_{ku}$ in a certain UoA $u$.
We assume that each publication $i$ has an intrinsic ``value'' $v_i$.
This value cannot be observed directly, but it does influence the peer review judgement that determines whether a publication is awarded four stars or not.
We assume that the value $v_i$ of publication $i$ is log-normally distributed as
\begin{equation}
  v_i \sim \lognormal(\mu_{ku}, 1).
  \label{eq:value_dist}
\end{equation}
The parameter $\mu_{ku}$ represents the capability of institution $k$ in UoA $u$ to produce publications of a high value.
The higher $\mu_{ku}$, the higher on average the values of publications of institution $k$ in UoA $u$.

Reviewers cannot directly observe the value $v_i$ of publication $i$.
Instead, they need to estimate this value.
When reviewers estimate the value $v_i$ of publication $i$, they may make a certain ``error''.
In other words, the value estimated by reviewers may differ from the true value.
We assume that the ``perceived value'' $p_i$ of publication $i$ is given by
\begin{equation}
  p_i = v_i \epsilon_i,
  \label{eq:error}
\end{equation}
where $\epsilon_i$ represents the error made in estimating the value of the publication.
We assume $\epsilon_i$ to be log-normally distributed as
\begin{equation}
  \epsilon_i \sim \lognormal\left(-\frac{\sigma^2_\epsilon}{2}, \sigma^2_\epsilon\right).
  \label{eq:error_dist}
\end{equation}
On average, $\epsilon_i$ equals $1$.
If $\epsilon_i > 1$, reviewers overestimate the value $v_i$.
On the other hand, if $\epsilon_i < 1$, they underestimate the value $v_i$.
However, on average, the perceived value $p_i$ equals the true value $v_i$. 

The inaccuracy of peer review is determined by the parameter $\sigma^2_\epsilon$.
If $\sigma^2_\epsilon = 0$, peer review is perfectly accurate, and the perceived value $p_i$ always equals the true value $v_i$.
Higher values of $\sigma^2_\epsilon$ correspond to less accurate peer review, so that the perceived value $p_i$ may differ from the true value $v_i$.

\begin{figure}%
	\includegraphics[width=\columnwidth]{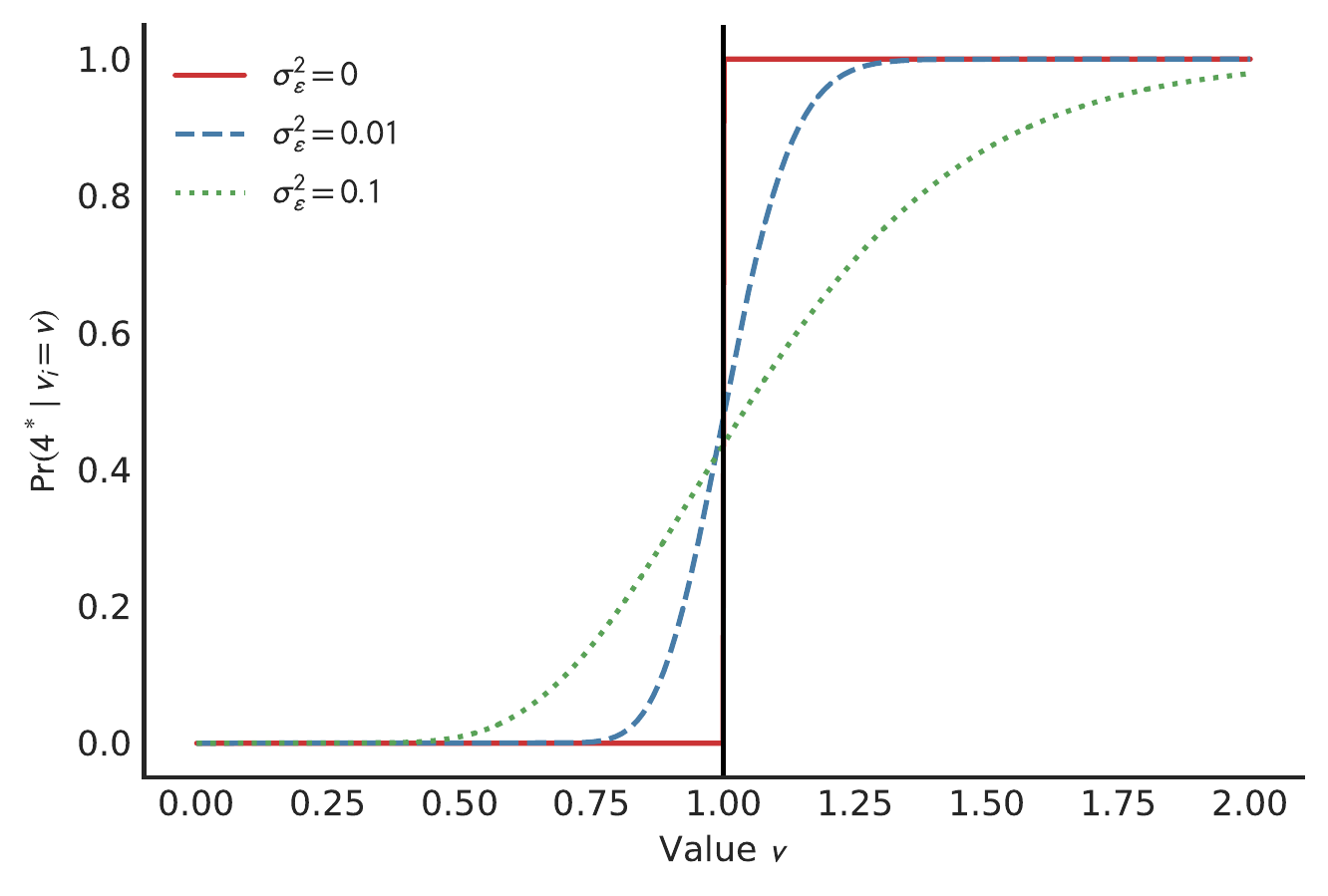}%
	\caption{Conditional probability that publication $i$ is awarded four stars given that it has value $v_i$.
	Three scenarios are considered for the inaccuracy of peer review $\sigma^2_\epsilon$.
	The threshold for being awarded four stars equals $p^{4^*} = 1$.}%
	\label{fig:prob}%
\end{figure}

Finally, we assume that publications with a perceived value higher than some threshold $p^{4^*}$ are awarded four stars.
Hence, publication $i$ is awarded four stars if $p_i > p^{4^*}$.
For any value $v_i$, there is some probability that publication $i$ is awarded four stars.
We denote this probability by $\Pr(i \text{~has~} 4^* \mid v_i)$.
For different values of $\sigma^2_\epsilon$, Fig.~\ref{fig:prob} shows how $\Pr(i \text{~has~} 4^* \mid v_i)$ depends on the value $v_i$.

\subsection{Parameters}

\noindent For each institution $k$ and UoA $u$, we estimate $\mu_{ku}$ based on the observed proportion of $4^*$ publications of institution $k$ in UoA $u$, denoted by $\PP_{ku}(4^*)$.
It follows from Eqs.~\ref{eq:value_dist}--\ref{eq:error_dist} that the perceived value $p_i$ is distributed as
\begin{equation}
  p_i \sim \lognormal\left(\mu_{ku} - \frac{\sigma^2_\epsilon}{2}, 1 + \sigma^2_\epsilon\right).
  \label{eq:perceived_value_dist}
\end{equation}
Denote by $F_{ku}$ the corresponding cumulative distribution.
Hence, $F_{ku}(p)$ equals the probability that $p_i \leq p$.
It follows that $1 - F_{ku}(p^{4^*})$ equals the probability that a publication is awarded four stars.
For a given $\sigma^2_\epsilon$ and $p^{4^*}$, the maximum likelihood estimate of $\mu_{ku}$ is obtained by choosing $\mu_{ku}$ such that $1 - F_{ku}(p^{4^*}) = \PP_{ku}(4^*)$.

Results obtained using our model seem to be largely independent of the threshold $p^{4^*}$.
We therefore simply set this threshold to $p^{4^*} = 1$.
We use values of $0.1$ and $1.0$ for $\sigma^2_\epsilon$, corresponding respectively to a relatively high and a relatively low accuracy of peer review.

\subsection{Estimating peer review uncertainty by resampling}

\begin{figure}%
	\includegraphics[width=\columnwidth]{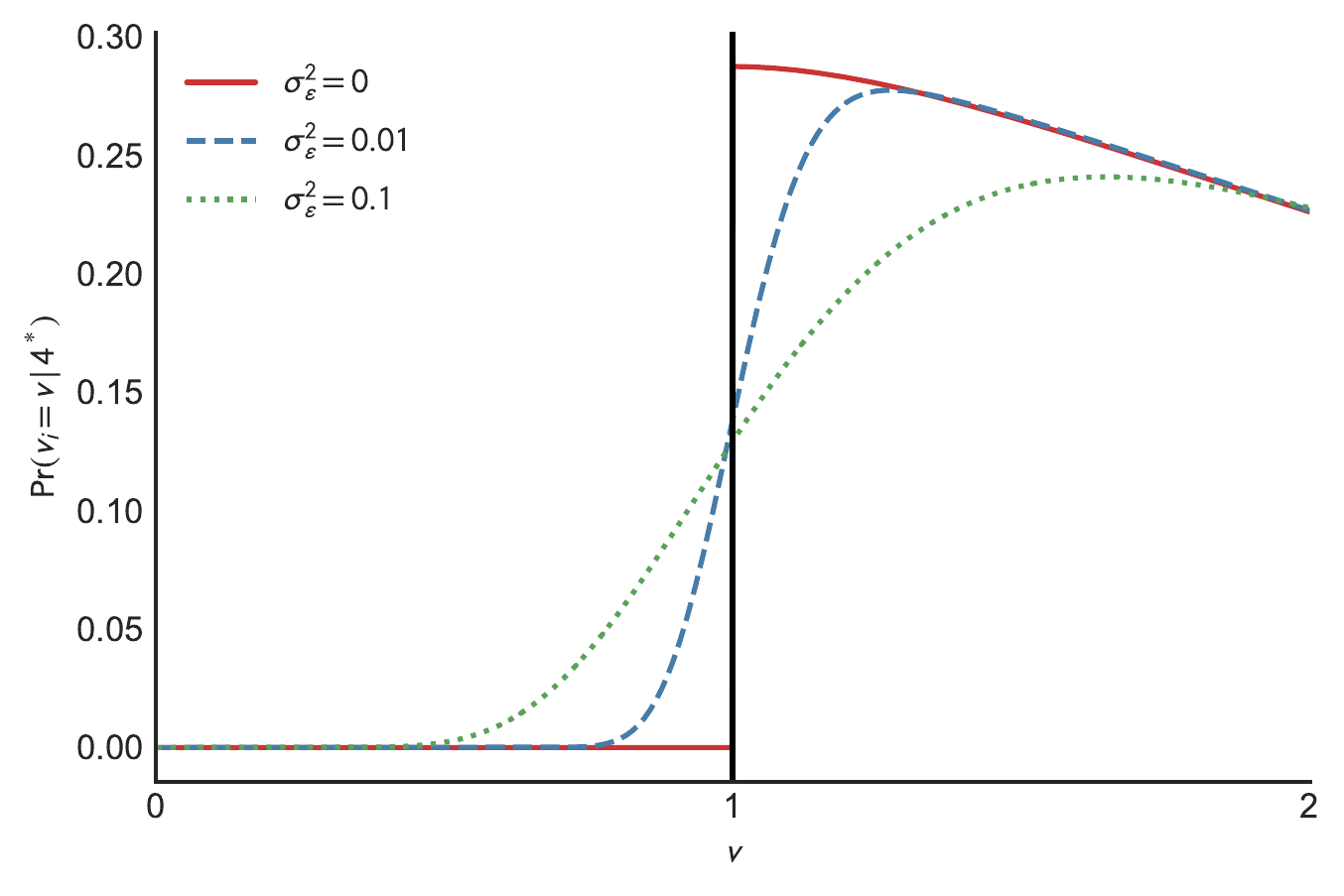}%
	\caption{Conditional probability that publication $i$ has value $v_i$ given that it was awarded four stars.
	Three scenarios are considered for the inaccuracy of peer review $\sigma^2_\epsilon$.
	The capability to produce high-value publications equals $\mu_{ku} = 1$.
	The threshold for being awarded four stars equals $p^{4^*} = 1$.}%
	\label{fig:cond_prob}%
\end{figure}

\noindent In the REF peer review results, we observe only whether a publication has been awarded four stars or not.
We do not observe the perceived value $p_i$ of a publication $i$, and clearly we do not know a publication's true value $v_i$.
To estimate the uncertainty of peer review in a certain field, we sample possible values for the publications of the institutions active in this field.
Using Bayes' theorem, the probability that publication $i$ has value $v_i$ given that it has four stars can be expressed as
\begin{align*}
	\Pr(v_i = v \mid i \text{~has~} 4^*) &= \Pr(v_i = v \mid p_i > p^{4^*}) \\
  &= \Pr(p_i > p^{4^*} \mid v_i = v) \frac{\Pr(v_i = v)}{\Pr(p_i > p^{4^*})} \\
	&= \Pr\left(\epsilon_i > \frac{p^{4^*}}{v}\right) \frac{\Pr(v_i = v)}{\Pr(p_i > p^{4^*})}.
  \label{eq:prob_cond_value}
\end{align*}
As illustrated in Fig.~\ref{fig:cond_prob}, the distribution of the value $v_i$ conditional on publication $i$ having four stars depends on the inaccuracy of peer review $\sigma^2_\epsilon$.
If peer review is perfectly accurate (i.e. $\sigma^2_\epsilon = 0$), we have $p_i = v_i$.
Consequently, if publication $i$ has four stars, we have $p_i > p^{4^*}$, which then implies $v_i > p^{4^*}$.
On the other hand, if peer review lacks perfect accuracy (i.e. $\sigma^2_\epsilon > 0$), some publications have four stars even though $v_i \leq p^{4^*}$.
These are publications for which reviewers have overestimated the value.
For these publications, the error $\epsilon_i$ is sufficiently large so that $p_i > p^{4^*}$ even though $v_i \leq p^{4^*}$.
Analogous considerations apply to $\Pr(v_i = v \mid i \text{~has no~} 4^*)$.

\begin{figure}%
	\includegraphics[width=\columnwidth]{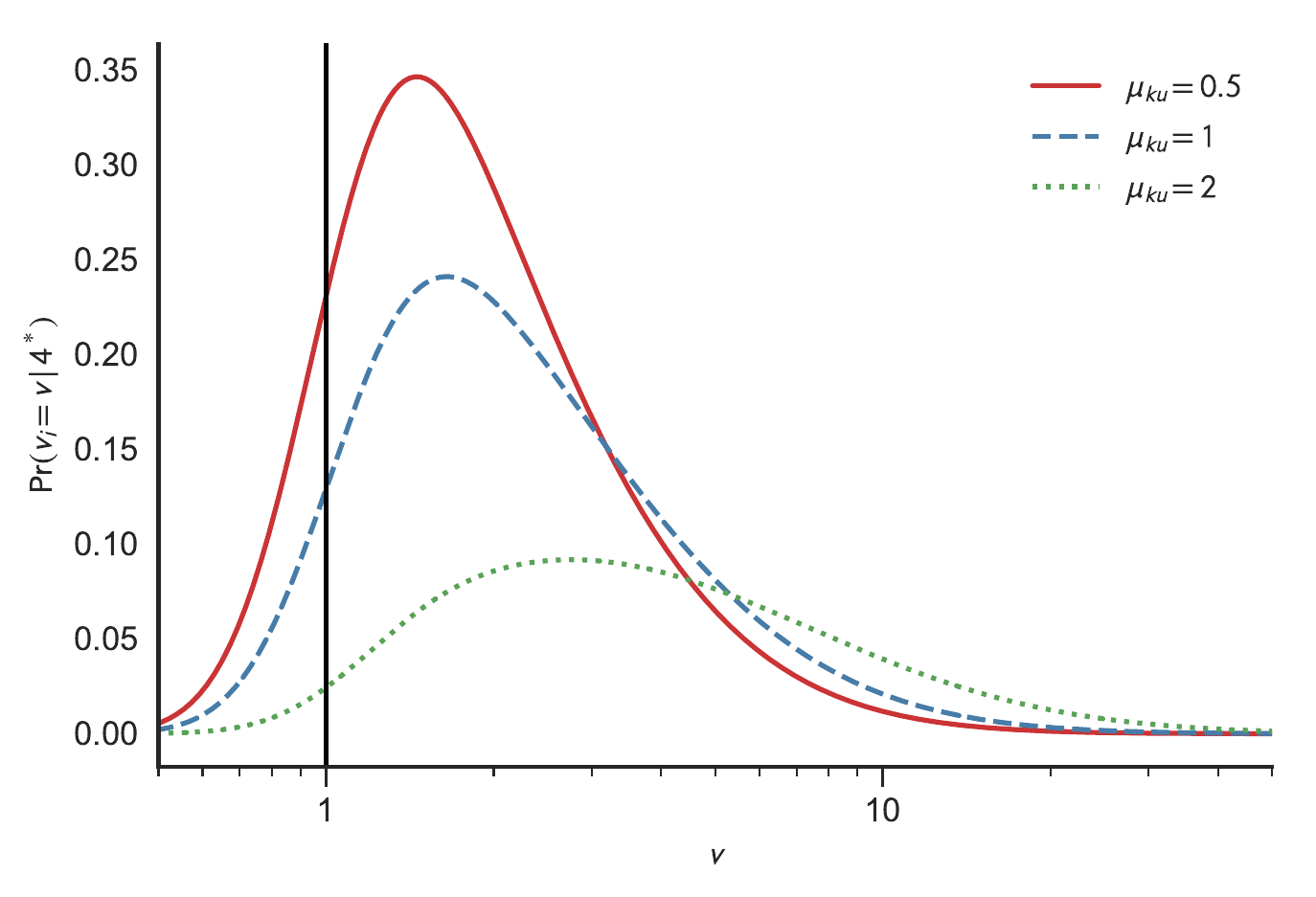}%
    \caption{Conditional probability that publication $i$ has value $v_i$ given that it was awarded four stars.
	Three scenarios are considered for the capability to produce high-value publications $\mu_{ku}$.
	The inaccuracy of peer review equals $\sigma^2_\epsilon = 0.1$.
	The threshold for being awarded four stars equals $p^{4^*} = 1$.}%
	\label{fig:cond_prob_mu}%
\end{figure}

For each institution $k$ in UoA $u$, we sample values $v_i'$ for all $n_{ku}$ publications.
For $4^*$ publications, we sample according to $\Pr(v_i' = v \mid i \text{~has~} 4^*)$.
For publications with fewer stars, we sample according to $\Pr(v_i' = v \mid i \text{~has no~} 4^*)$.
For institutions with higher $\mu_{ku}$, we are more likely to sample higher values, as illustrated in Fig.~\ref{fig:cond_prob_mu}. 
Next, for each publication $i$, we sample a perceived value $p_i'$.
This perceived value is given by $p_i' = v_i' \epsilon_i'$, where the error $\epsilon_i'$ is sampled according to Eq.~\ref{eq:error_dist}.
We then calculate the proportion of publications for which $p_i' > p^{4^*}$.
This yields the proportion of publications of institution $k$ in UoA $u$ that would be awarded four stars based on the sampled perceived values.
We denote this proportion by $\PP'_{ku}(4^*)$.

Finally, for each field $u$, we calculate the agreement between $\PP_{ku}(4^*)$ and $\PP'_{ku}(4^*)$ for all institutions $k$.
This offers an indication of the uncertainty of peer review in field $u$.
For the results presented in Section~\ref{sec:results}, we resample $1\,000$ times.
We report both the median outcome obtained from $1\,000$ samples and the interval that covers $95\%$ of the outcomes.

\FloatBarrier

\section{Detailed results}

{
\rowcolors{2}{Set1-blue!20}{white}
\setlength{\tabcolsep}{8pt}
\LTcapwidth=\textwidth

\begin{longtable}{lp{8cm}rrrr}
\caption{Agreement per unit of assessment.}\label{tab:agreement}\\
\toprule
\rowcolor{white}
  &                                                                  & \multicolumn{2}{l}{Size-independent} & \multicolumn{2}{l}{Size-dependent} \\
\rowcolor{white}
  & Unit of assessment  &                R &  MAD &              R &  MAPD \\
\midrule
\endfirsthead
\toprule
\rowcolor{white}
  &                                                                  & \multicolumn{2}{l}{Size-independent} & \multicolumn{2}{l}{Size-dependent} \\
\rowcolor{white}
  & Unit of assessment  &                R &  MAD &              R &  MAPD \\
\midrule
\endhead
\midrule
\rowcolor{white}
\multicolumn{3}{r}{{Continued on next page}} \\
\midrule
\endfoot

\bottomrule
\endlastfoot
1  & Clinical Medicine                                                             & 0.86 & 2.9  & 0.98 & 14.7 \\
2  & Public Health, Health Services and Primary Care                               & 0.83 & 3.0  & 0.98 & 15.8 \\
3  & Allied Health Professions, Dentistry, Nursing and Pharmacy                    & 0.54 & 4.8  & 0.95 & 25.6 \\
4  & Psychology, Psychiatry and Neuroscience                                       & 0.79 & 4.0  & 0.96 & 27.1 \\
5  & Biological Sciences                                                           & 0.69 & 9.0  & 0.98 & 31.8 \\
6  & Agriculture, Veterinary and Food Science                                      & 0.68 & 4.8  & 0.96 & 27.1 \\
7  & Earth Systems and Environmental Sciences                                      & 0.80 & 3.1  & 0.93 & 25.2 \\
8  & Chemistry                                                                     & 0.83 & 4.3  & 0.94 & 23.2 \\
9  & Physics                                                                       & 0.86 & 2.8  & 0.98 & 12.1 \\
10 & Mathematical Sciences                                                         & 0.39 & 5.9  & 0.96 & 33.4 \\
11 & Computer Science and Informatics                                              & 0.49 & 7.3  & 0.87 & 46.1 \\
12 & Aeronautical, Mechanical, Chemical and Manufacturing Engineering              & 0.39 & 4.1  & 0.98 & 29.8 \\
13 & Electrical and Electronic Engineering, Metallurgy and Materials               & 0.77 & 4.9  & 0.91 & 35.9 \\
14 & Civil and Construction Engineering                                            & 0.28 & 5.1  & 0.87 & 28.7 \\
15 & General Engineering                                                           & 0.51 & 5.2  & 0.90 & 43.0 \\
16 & Architecture, Built Environment and Planning                                  & 0.29 & 10.6 & 0.83 & 78.5 \\
17 & Geography, Environmental Studies and Archaeology                              & 0.68 & 8.0  & 0.85 & 45.1 \\
18 & Economics and Econometrics                                                    & 0.89 & 5.9  & 0.95 & 27.4 \\
19 & Business and Management Studies                                               & 0.60 & 4.4  & 0.96 & 33.7 \\
20 & Law                                                                           & 0.40 & 8.1  & 0.87 & 85.2 \\
21 & Politics and International Studies                                            & 0.52 & 5.8  & 0.90 & 38.9 \\
22 & Social Work and Social Policy                                                 & 0.36 & 7.3  & 0.88 & 44.6 \\
23 & Sociology                                                                     & 0.27 & 4.8  & 0.86 & 25.9 \\
24 & Anthropology and Development Studies                                          & 0.10 & 8.3  & 0.63 & 42.3 \\
25 & Education                                                                     & 0.41 & 8.4  & 0.95 & 45.9 \\
26 & Sport and Exercise Sciences, Leisure and Tourism                              & 0.35 & 7.7  & 0.84 & 43.8 \\
27 & Area Studies                                                                  & 0.10 & 12.8 & 0.48 & 61.3 \\
28 & Modern Languages and Linguistics                                              & 0.35 & 15.9 & 0.58 & 80.2 \\
29 & English Language and Literature                                               & 0.34 & 14.2 & 0.56 & 62.2 \\
30 & History                                                                       & 0.40 & 9.9  & 0.90 & 56.1 \\
31 & Classics                                                                      & 0.21 & 16.8 & 0.49 & 63.3 \\
32 & Philosophy                                                                    & 0.42 & 13.6 & 0.83 & 52.3 \\
33 & Theology and Religious Studies                                                & 0.35 & 17.5 & 0.62 & 87.9 \\
34 & Art and Design: History, Practice and Theory                                  & 0.39 & 14.0 & 0.56 & 98.7 \\
35 & Music, Drama, Dance and Performing Arts                                       & 0.33 & 17.8 & 0.55 & 97.6 \\
36 & Communication, Cultural and Media Studies, Library and Information Management & 0.42 & 12.9 & 0.68 & 72.4 \\
\end{longtable}

}

\FloatBarrier

\begin{figure*}[t]
  \centering
  \includegraphics[width=\linewidth]{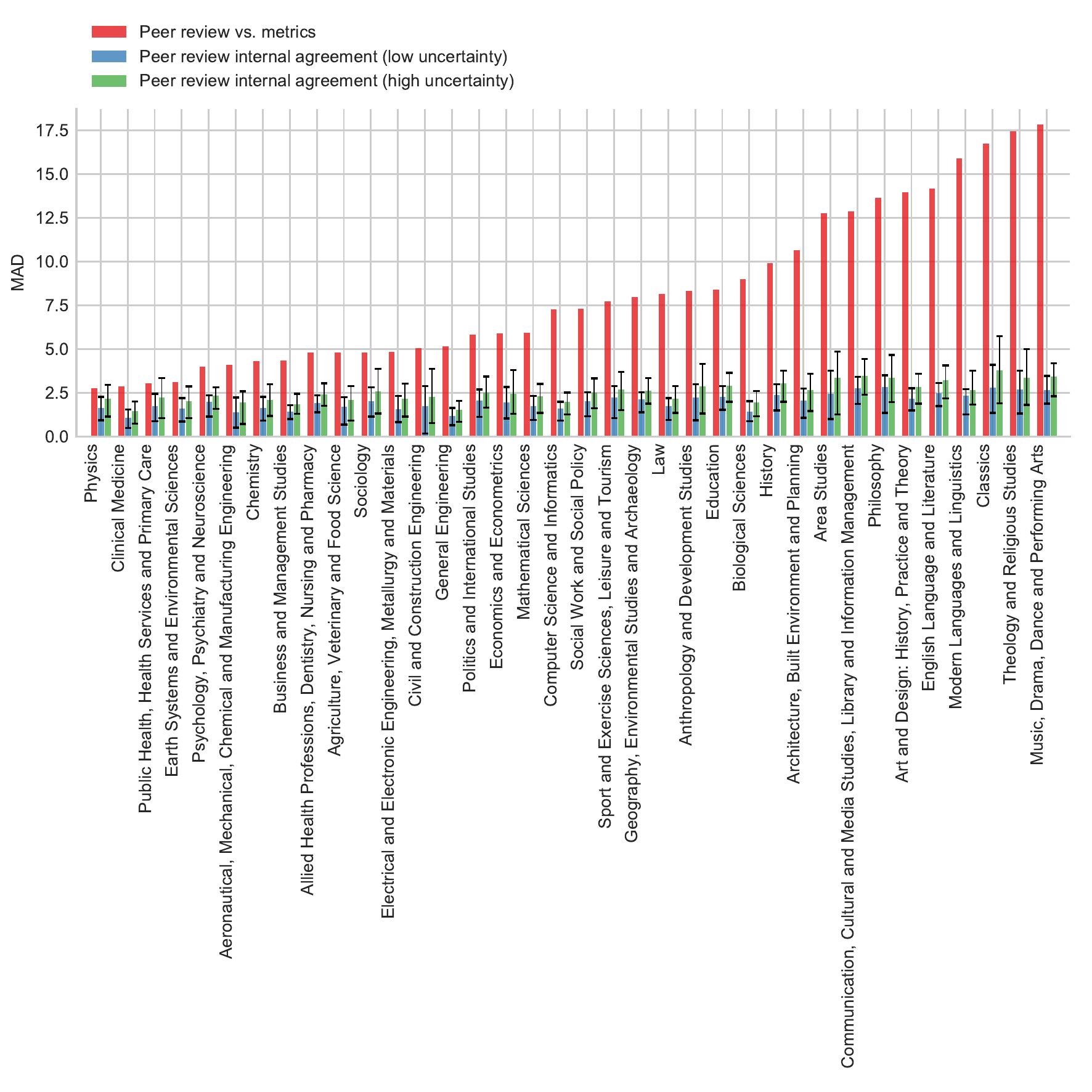}
  \caption{Size-independent median absolute difference (MAD) between $\PP(\text{top}~10\%)$ and $\PP(4^*)$ compared with the MAD based on a model of peer review uncertainty for all units of assessment.}
  \label{fig:MAD_barchart_size_indep_all}
\end{figure*}

\begin{figure*}[t]
  \centering
  \includegraphics[width=\linewidth]{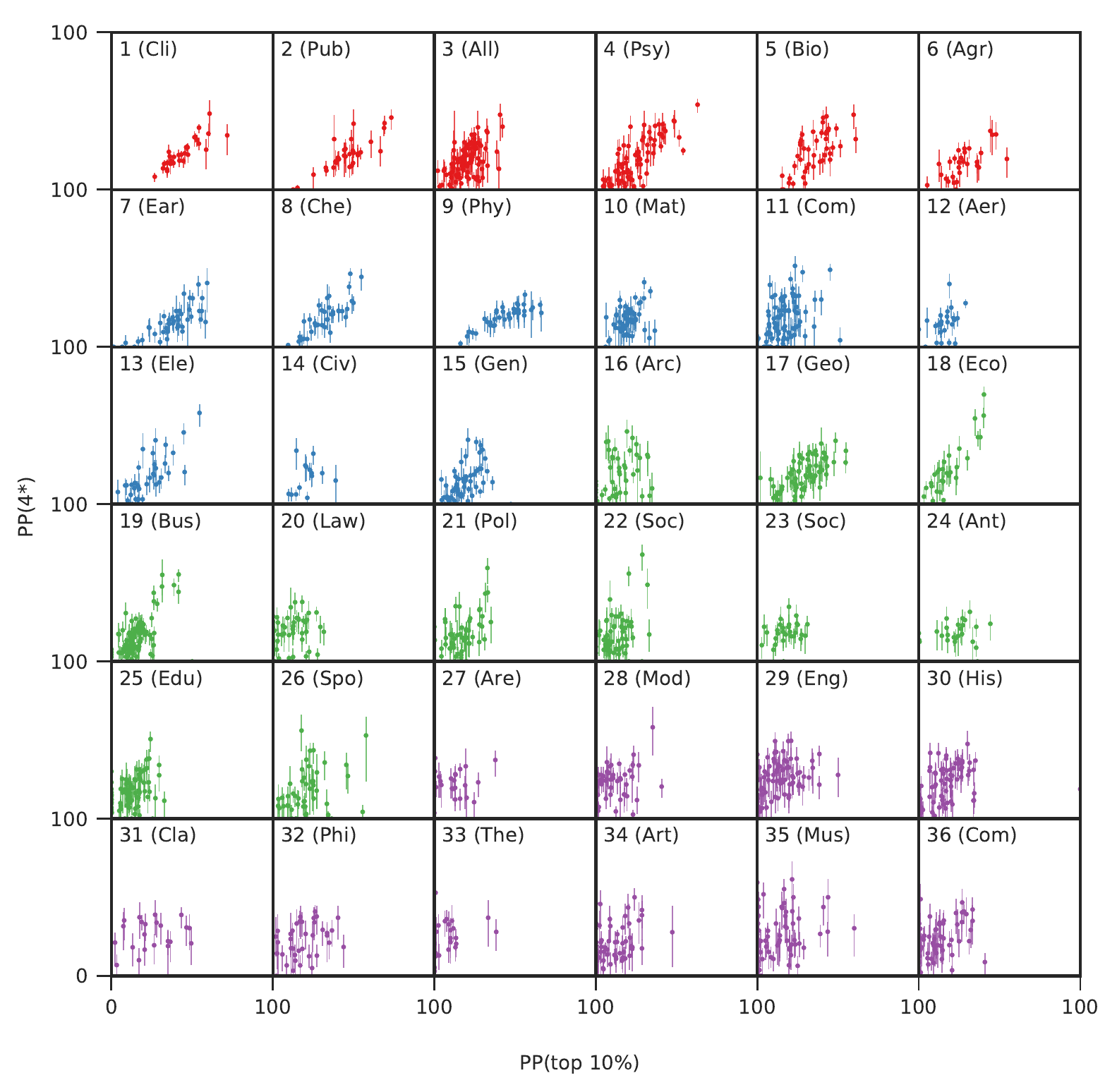}
  \caption{Scatter plots of $\PP(\text{top}~10\%)$ and $\PP(4^*)$ at the institutional level for all units of assessment. 
  Error bars indicate the $95\%$ interval of bootstrapped peer review results for low peer review uncertainty.
  }
  \label{fig:scatter_size_indep}
\end{figure*}

\begin{figure*}[t]
  \centering
  \includegraphics[width=\linewidth]{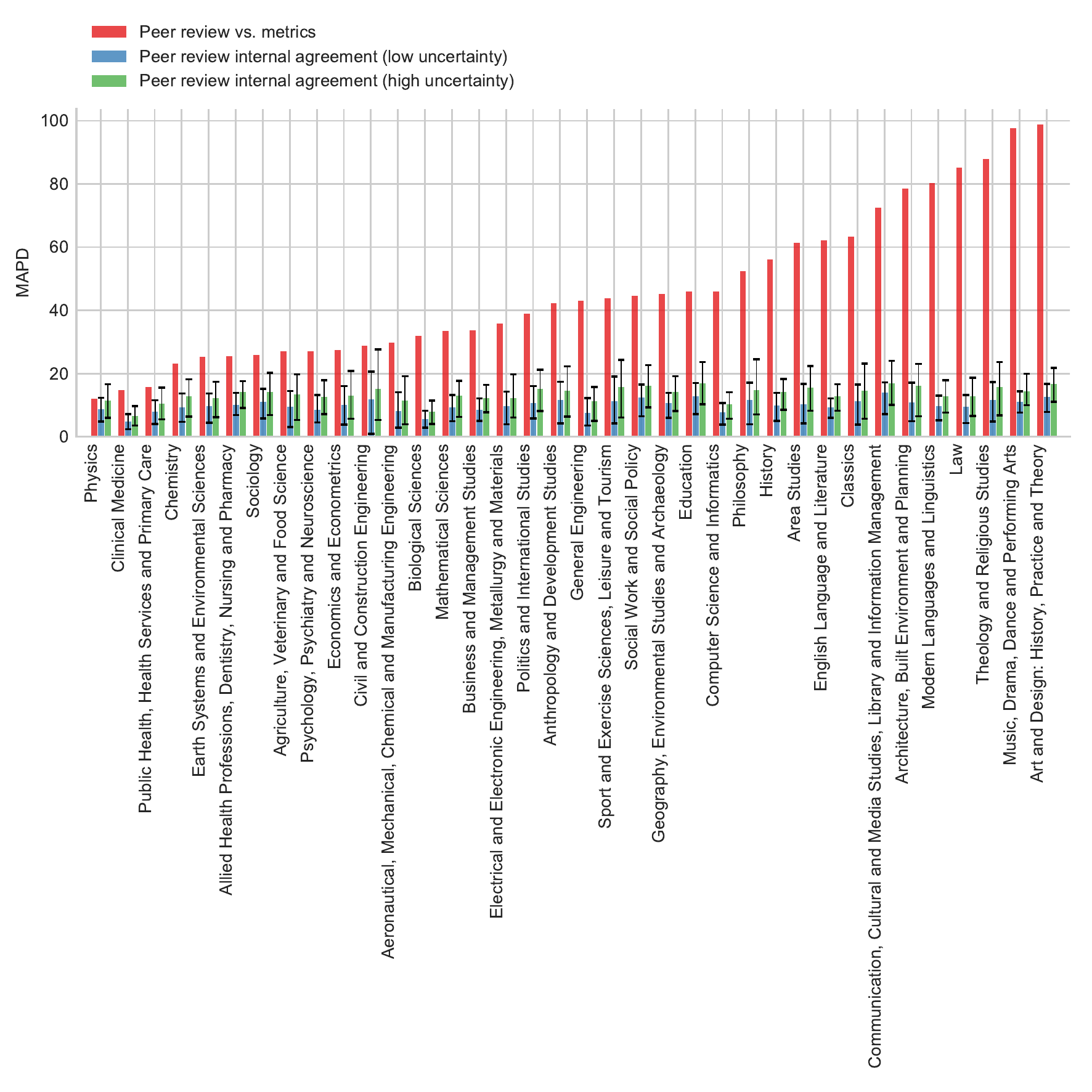}
  \caption{Size-dependent median absolute percentage difference (MAPD) of $\Pub(\text{top}~10\%)$ relative to $\Pub(4^*)$ compared with the MAPD based on a model of peer review uncertainty for all units of assessment.}
  \label{fig:MAPD_barchart_size_dep_all}
\end{figure*}

\begin{figure*}[t]
  \centering
  \includegraphics[width=\linewidth]{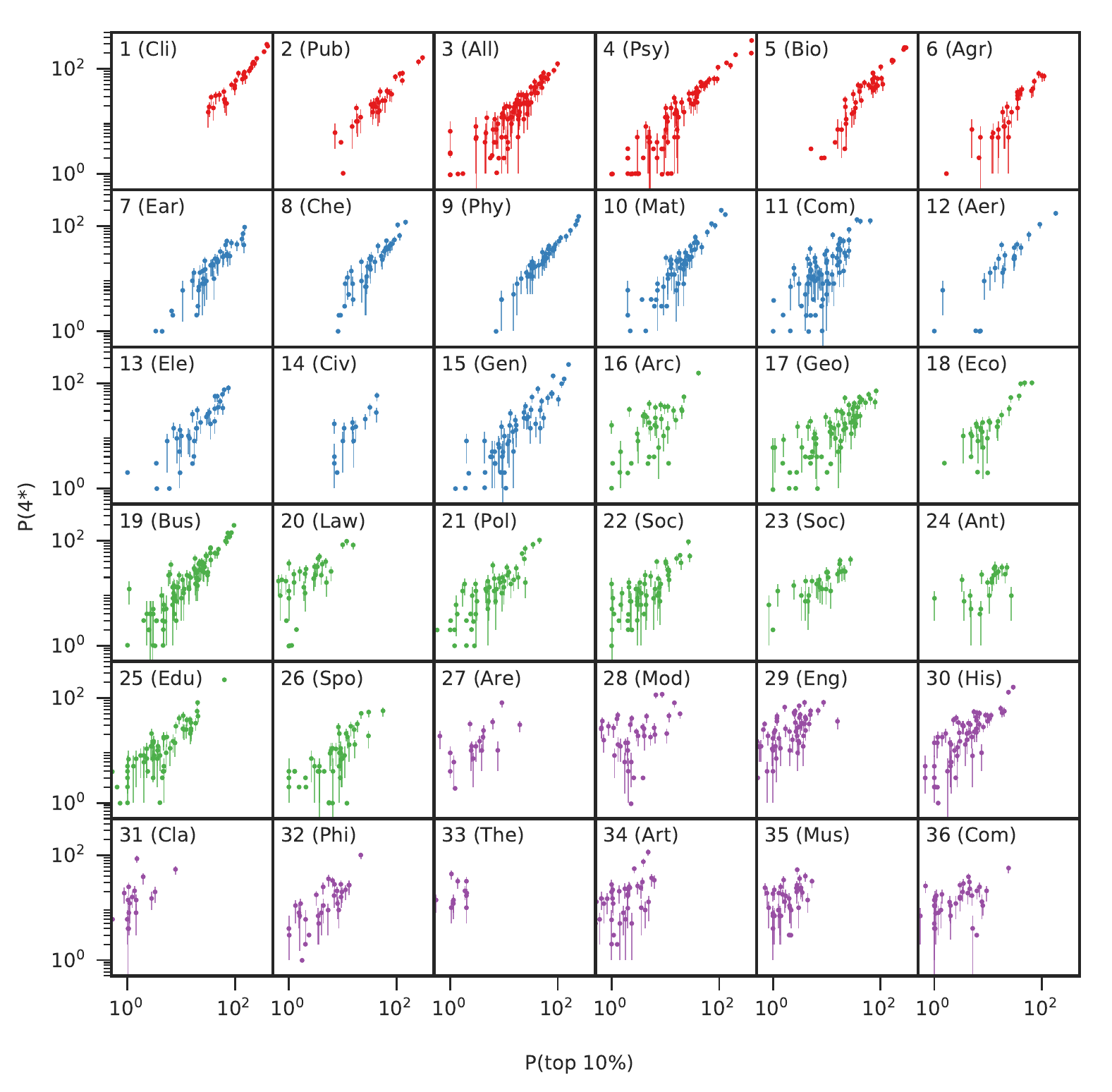}
  \caption{Logarithmic scatter plots of $\Pub(\text{top}~10\%)$ and $\Pub(4^*)$ at the institutional level for all units of assessment.
   Error bars indicate the $95\%$ interval of bootstrapped peer review results for low peer review uncertainty.
   }
  \label{fig:scatter_size_dep}
\end{figure*}

\end{document}